\documentclass[]{elsarticle}
\usepackage{booktabs} 
\usepackage{tabularx}
\usepackage[margin=1.2in]{geometry}
\usepackage{natbib}
\usepackage{lscape}

\usepackage{listings}
\usepackage{epstopdf}
\usepackage{multirow, makecell}
\usepackage{subfig}
\usepackage{graphicx}

\usepackage{hyperref}
\hypersetup{
    colorlinks=true,
    linkcolor=blue,
    filecolor=magenta,      
    urlcolor=cyan,
}

\makeatletter
\def\ps@pprintTitle{%
  \let\@oddhead\@empty
  \let\@evenhead\@empty
  \let\@oddfoot\@empty
  \let\@evenfoot\@oddfoot
}
\makeatother

\newcolumntype{Y}{>{\centering\arraybackslash}X}

\journal{Elsevier}

\usepackage{numcompress}\biboptions{authoryear} 

\begin{document}












\author
{Kalle Koivisto$^{1\ast}$ and Toni Taipalus $^{2}$
\normalsize{$^{1}$University of Jyväskylä, Finland}
\normalsize{$^{2}$Tampere University, Finland}
\normalsize{$^\ast$Corresponding author: Toni Taipalus, E-mail: toni.taipalus@tuni.fi}
\footnote{This is a pre-print version of the article. Please cite the publisher's version of the article: Koivisto, K. \& Taipalus, T. (2025). Pitfalls in Effective Knowledge Management: Insights from an International Information Technology Organization. \textit{International Journal of Knowledge Management Studies} 16(1), DOI: \url{10.1504/IJKMS.2025.146083}.}
}

\begin{frontmatter}

\title{Pitfalls in Effective Knowledge Management:\\ Insights from an International Information Technology Organization}



\begin{abstract}
Knowledge is vital for organizations, but effective knowledge management remains a challenge. Despite the acknowledged importance of knowledge sharing and management, many organizations struggle to harness their knowledge effectively, leading to cooperation issues and the loss of valuable insights when employees leave. This study, based on interviews with 50 employees in a large international IT company, aims to identify and address hindering factors that impede knowledge sharing and management. It reveals a significant gap between the perceived importance of knowledge management and its practical implementation. The study identifies 44 hindering factors in various categories. Recommendations for mitigating these obstacles include offering training and guidelines to improve employee actions. These findings benefit knowledge-intensive organizations by informing strategies to enhance performance.
\end{abstract}


\end{frontmatter}

\section{Introduction}
\label{sec-introduction}

Organizations' most important strategic resource in today's highly competitive environment is knowledge and the ability to transfer it well \citep{Ruzic_2021}. Managing knowledge effectively enables organizations to avoid mistakes and redundancy \citep{Ahmed_2006}, increase organizational performance and effectiveness, strengthen their ability to match customer needs \citep{Kim_2006}, and gain competitive advantage \citep{Hume_2015,Braganza_2004}. Managing knowledge effectively is a prerequisite to competing in today's markets and should thus be carefully studied and understood both by practitioners and researchers.

As a knowledge-intensive intellectual activity, software development is one of the fields where knowledge is one of the most important resources influencing both the success and performance of IT teams and organizations \citep{Ryan_2009,Ryan_2013}. Knowledge management literature has emphasized that it is knowledge sharing - a subset of knowledge management - that enables organizations to develop a competitive advantage and enhance employees' capacity to innovate creative solutions \citep{Jackson_2006}. In fact, it is the individual's knowledge-sharing behavior that helps organizations to make quick decisions that are effective and resilient during crises to keep the organization operational at all times \citep{Wang_2010}. Considering that the individual's choice and behavior with knowledge can define a big part of how well an organization is performing, researchers should focus their efforts in understanding the reasons behind these choices and behavior and whether they are promoting knowledge sharing or hindering it.

Prior research has mostly focused on investigating the most common challenges in knowledge management and best practices to meet these challenges. These existing challenges and practices are widely documented and explained but it seems that much less attention, if at all, has been paid to understanding why these practices are not being implemented, integrated, and used effectively. Prior research confirms that software development organizations and their employees do perceive the benefits and importance of knowledge management and sharing, yet their efforts to share and manage knowledge effectively are poorly planned, inconsistent, and often of poor quality \citep{Dingsoyr_2009,Aurum_2008,Prikladnicki_2003}. This study argues that factors preventing and hindering effective knowledge management need to be identified so that this phenomenon can properly be addressed.

The aim of this study is to address the discovered gap through an interpretive case study where the data is collected by conducting semi-structured group interviews. 22 teams in total were included to capture data from all the possible functions within the case organization. The goal is to discover, identify, and understand any possible factors that might hinder or prevent the adoption of already documented knowledge management practices and thus effective knowledge management. This study focuses on knowledge being shared and managed within and between teams related to software development. This study also attempts to provide a few recommendations on how to address the identified hindering factors. These two objectives are reflected in the research questions this study attempts to answer:

\begin{itemize}
\item[\textbf{RQ1}] Which factors hinder and complicate knowledge sharing and knowledge management within and between globally distributed cross-functional teams?
\item[\textbf{RQ2}] What kind of measures can be taken to solve hindering factors in knowledge sharing and knowledge management within and between globally distributed cross-functional teams?
\end{itemize}

The results indicate that the 44 hindering factors for effective knowledge management can be divided into five partially overlapping themes or topics: personal social, organizational social, technical, environmental, and interrelated social and technical. Our five recommendations drawn from these factors emphasize planning, training, focus, guidelines, and document labeling to mitigate the identified factors that hinder effective knowledge management.

The rest of the study is structured as follows. The next section acquaints the reader with knowledge management, and known challenges and practices in knowledge management. Next, in Section 3, the methodology of this study and data collection is presented. The results are presented in Section 4 and discussed in Section 5. Section 6 concludes the study.

\section{Background and related work}
\label{sec-background}

\subsection{Knowledge management}
\label{sec-bg-knowledge-management}

Knowledge management process involves capturing, storing, and sharing knowledge within an organization to improve decision-making and promote continuous learning and improvement. Effective knowledge management is vital for software development organizations, as knowledge is the main strategic resource that requires protection and successful transfer \citep{Ruzic_2021,Rus_2002}. Benefits of effective transfer include avoiding mistakes and redundancy, strengthening entrepreneurial orientation, gaining competitive advantage \citep{Ode_2020}, reducing time to market, improving organizational effectiveness and performance, and enhancing the ability to meet customer needs \citep{Ahmed_2006,De_Clercq_2013,Hume_2015,Kim_2006}. However, knowledge needs to be effectively managed to provide any benefits to companies \citep{Andreeva_2012}. Both \cite{Choi_2010} and \cite{Oliva_2019} found that the mere existence of knowledge or even knowledge sharing does not positively affect team performance unless it is effectively applied. Effective knowledge management consists of knowledge processes (creation, acquisition, transfer, sharing, and application) and management activities that support these processes \citep{Gold_2001,Lee_2003}. In order to draw value from and improve these processes, effective knowledge management practices must be in place \citep{Andreeva_2012}.

Software development is typically performed in teams, and team members learn from each other's experiences and projects. If this new knowledge is not shared, the team and organization miss an opportunity to benefit from it \citep{Rus_2002}. Effective coordination of this knowledge is crucial, especially for knowledge workers like software developers who require special skills and expertise \citep{Faraj_2000,Ryan_2009}. Knowledge sharing within teams is particularly important, as knowledge from colleagues is more likely to be relevant and attainable \citep{Aurum_2008}. Knowledge sharing is a subset of knowledge management that enhances employees' capabilities and helps create a competitive advantage \citep{De_Gooijer_2000,Jackson_2006,Al_2022}. Effective knowledge sharing between software development teams is a crucial success factor, as these teams must work in a cross-functional environment \citep{Aurum_2008}.

\cite{Bartol_2002} define knowledge sharing as the process of transferring explicit knowledge to other members of an organization. Sharing any knowledge that can be reasonably shared with another person is considered valuable, even if it is tacit knowledge that can be codified into explicit knowledge. Team members play the most influential role in knowledge sharing within and across teams, although managers can have an enabling effect on this matter by improving knowledge sharing between team members, teams, and the entire organization \citep{Bailey_2001,Abubakar_2019}. \cite{Lin_2008} identifies three factors that affect knowledge sharing: organizational structure, including complexity and employee centralization; organizational culture, including supportive, innovative, creative, and bureaucratic cultures; and the interaction between departments. Findings of \cite{Singh_2021} support the aspect of innovation. \cite{Abili_2011} suggest that effective knowledge sharing occurs when employees have the necessary power, as centralization in the organization prevents effective knowledge sharing.

Employee turnover in the IT industry is a common challenge that organizations face, where skilled professionals leave and take with them valuable skills, experience, and knowledge \citep{Rus_2002}. This creates a gap in the organization that needs to be filled, leading to loss of productivity as new hires need time to learn and adapt to the organization \citep{Battin_2001}. Organizational challenges such as inadequate metrics to measure the success of knowledge management initiatives, and the lack of effective knowledge management culture can also hinder knowledge sharing and management \citep{Lee_2003,Aurum_2008,Butt_2019}. The size of the organization can also impact the effectiveness of knowledge management, with larger organizations experiencing structural inertia and limitations in knowledge sharing due to breakdowns in communication channels \citep{Fores_2016,Connelly_2003}.

Agile methodologies used in global software development have been shown to encourage team members to share knowledge by interacting with each other directly rather than documenting knowledge, leading to uneven concentrations of knowledge and outdated documentation in repositories \citep{Beck_2001,Manteli_2011}. Social relationships between colleagues are crucial in facilitating knowledge transfer, with strong relationships leading to positive knowledge-sharing behavior \citep{Wendling_2013,Alavi_2002}. Distributed software development can make it harder to organize social interactions, leading to weaker relationships between colleagues, which can hinder knowledge transfer \citep{Szulanski_1996,Asp_2021}. Remote colleagues may also be perceived as threats, leading to resistance and fear among employees \citep{Nahapiet_1998}.

\subsection{Challenges in knowledge management}
\label{sec-bg-challenges}

Global software development is a complex process that involves managing and sharing knowledge across different locations. The challenges of knowledge management and sharing have been acknowledged since the adoption of globally distributed software development \citep{Herbsleb_2001}. To successfully utilize a globally distributed workforce, organizations must carefully consider knowledge-sharing challenges \citep{Wendling_2013}.

Communication challenges arise due to the separation of communication into local and remote channels. Local communication can be face-to-face, whereas remote communication relies on information and communication technology (ICT) systems \citep{Agerfalk_2005}. Geographical distance negatively affects team knowledge sharing, as information exchange during informal local interactions is not distributed to remote team members \citep{Taweel_2009,Asp_2021}. Internal communication is a priority for productivity \citep{Mazzei_2010}, and cultural and language differences affect communication efficiency \citep{Bolisani_2023,Kalkan_2008}. Face-to-face interaction enables non-verbal communication, which clarifies communication between colleagues with different cultural and lingual backgrounds \citep{Klitmoller_2013,Zieba_2023}. Lack of remote communication can lead to weakened communication within or between teams, which can threaten the benefits of distributed working models \citep{Herbsleb_2001b}. Knowing whom to contact and how to reach certain individuals in time can be difficult, especially for novices \citep{Desouza_2006}, leading to issues with social relationships and mutual trust.

Documentation challenges arise from the need to transform tacit knowledge into explicit knowledge that can be accessed by other employees of the organization \citep{Almeida_2014}. Explicit knowledge can be presented in different kinds of documents through the codification of knowledge \citep{Gunnlaugsdottir_2003,Razzaq_2019}. Virtual teams are strongly dependent on explicit knowledge \citep{Griffith_2003}, and it should be documented and stored where it can be (re)used and accessed by other employees. The ''people-to-document`` approach enables other employees to search and retrieve documented, codified knowledge themselves and not rely on the person who originally stored the information \citep{Hansen_1999}. A centralized platform or repository is needed to search, find, and retrieve knowledge for further use \citep{Almeida_2014}. A knowledge management system (KMS) enables the documentation, distribution, and transfer of explicit and tacit knowledge between employees in different ways \citep{Voelpel_2005,Noe_2003}. The ISO 30401:2018 Standard \citep{iso30401} provides guidelines and best practices for the establishment, implementation, maintenance, and improvement of a KMS, aiming to enhance knowledge sharing, collaboration, and organizational performance. The standard focuses on various aspects of knowledge management, including knowledge creation, capture, storage, transfer, and measurement. It promotes the effective management of intellectual capital and supports organizations in leveraging their knowledge assets for sustainable success. KMSs provide a way for the organization to better utilize internal knowledge resources and transfer that into competitive advantage. However, updating the already existing knowledge can be seen as difficult, and not given high priority by the organization \citep{Aurum_2008}. This can lead to lower quality content and trust issues in the KMS \citep{Markus_2001,Machado_2022}, which can discourage individuals, teams, or the organization as a whole from sharing knowledge with others. 

In summary, it has been noted that knowledge management issues are complex, and causes of realized problems can often only afterwards be traced back to knowledge management issues \cite{Nakash_2022}. As both knowledge management issues and benefit are indirect and delayed, it has been shown that knowledge management investments can be overlooked, and difficult to understand and justify \citep{Nakash_2023,Trkman_2012}.

\subsection{Knowledge sharing practices}
\label{sec-bg-practices}

A common argument about agile methodologies is that it prohibits documentation. Agile methodologies place less emphasis on documentation, but documentation is still created and maintained when the team sees it as suitable \citep{Dorairaj_2012}. Despite this guideline, adopting agile methodologies can also bring knowledge-sharing benefits to organizations, and thus following agile methodologies are seen as a known knowledge-sharing practice. In the Manifesto for Agile Software Development, the importance of interactions and individuals is emphasized \citep{Beck_2001} which would speak in favor of local teams when it comes to software development. \cite{Ryan_2009} found that social interactions were an excellent way to share tacit knowledge, which in turn positively influences team performance \citep[cf. e.g.,][]{Nonaka_1995}.

In distributed teams, communication must be done mostly through technological means. One of these communication channels is audio and video conferencing. Studies have noticed that we prefer textual communication over audio when we deal with simple, repetitive matters, but are more likely to have audio involved when we are discussing more complex topics \citep{Niinimaki_2010}. However, using audio or video conferencing tools in one-to-one discussions can be perceived as intrusive, and thus quite in-efficient unless the case is urgent and necessary to deal between two people.

Another widely used communication channel that has already been used for years is email and mailing lists. Compared to other means of communication, email has a much more formal appearance and nature. Email can also be used as a permanent storage location for different documents and communication \citep{Manteli_2011,Niinimaki_2010}.
 
Tacit knowledge is most likely efficiently gained through experiences \citep{Polanyi_1966,Sternberg_2000}. These experiences are not easy to gain and thus one of the best ways to do so is to learn by performing various daily activities \citep{Sternberg_2000}. \cite{Aurum_2008} detected that practitioners also recognize the learning-by-doing method as one of the most common sources of knowledge. As this has been identified as one of the best methods to acquire knowledge \citep{Zaim_2019}, several practices have been invented to replicate this learning by doing. To simulate learning by doing organizations are encouraging practitioners to seek novel assignments, be mentored by experienced colleagues, assign specific tasks, and enable job rotation \citep{Sternberg_2000}. 

\cite{Alavi_2001} proposed that the application of ICT can create an environment and infrastructure that contributes to knowledge management by supporting and augmenting a variety of knowledge processes. The concept of a KMS is widely mentioned in different knowledge-related studies and is considered a basic tool for organizations to enhance their knowledge management processes \citep{Alavi_2002,Aurum_2008,Dingsoyr_2013,Dorairaj_2012,Prikladnicki_2003,Taweel_2009}. IT organizations seem to prefer wiki-based knowledge management systems where knowledge can be stored in an organized way with comprehensive search capability and document versioning \citep{Dorairaj_2012,Taweel_2009}.

Employee turnover is a common challenge that many organizations face especially in the IT sector. When experienced workers or experts leave the company, they are often replaced by individuals that have less experience and a lot of knowledge to be acquired especially at the start of the new employment \citep{Rus_2002}. The start of the new employment very often contains an onboarding process where the new employee is given the necessary skills and knowledge to be able to become a productive member of the organization. Documented knowledge, usually in some sort of KMS, is also widely used and highly beneficial for new employees because they are able to access a lot of information at once \citep{Dorairaj_2012,Taweel_2009} from a location that is perceived as reliable \citep{Desouza_2006}. 

Organizations' management plays a crucial role in the success of organization-wide knowledge management practices. Several studies have reported that current knowledge management practices are not sufficient. \cite{Prikladnicki_2003} studied two organizations that were both missing a consistent and formal knowledge management process causing notable challenges in knowledge sharing. \cite{Dingsoyr_2009} conducted a survey where the current situation of knowledge management was compared to the targeted future situation and found that practitioners see a lot of room for improvements in knowledge management practices. \cite{Witherspoon_2013} suggested that knowledge sharing should be built upon attitudes and intentions, rewards, and organizational culture, but these views have also received criticism \citep{Hau_2013,Bock_2002,Seba_2012}. Based on \cite{Szulanski_1996}, attitudes and intentions toward knowledge sharing can be positively affected by routines and guidance that lower the perceived difficulty of the knowledge-sharing process. Lowering the perceived difficulty is key because it has been found that knowledge sharing within organizations is dependent on individuals' behaviour \citep{Bock_2005} and their ability to share knowledge \citep{Gressgard_2015}. Whether it's a dedicated role in the team responsible for knowledge management or the responsibility is shared evenly with all the team members to manage their own knowledge, it is pivotal that the responsibility is clearly stated and understood by all involved \citep{Aurum_2008}. 

The same social practices promoting local knowledge sharing, like informal discussions with colleagues, are the ones that make global knowledge management challenging \citep{Desouza_2006}. \cite{Kotlarsky_2005} studied if social ties and knowledge sharing in global software development teams support collaboration. Social ties referred to trust and rapport whereas knowledge sharing referred to collective knowledge and transactive memory. Not much attention was paid to the social aspects of globally distributed collaboration or its influence on coordination when the study was initially conducted. Most of the existing solutions at the time were technical by nature rather than social. Researchers claimed that collaboration could be improved in globally distributed software development teams through knowledge sharing and social ties. They suggested that organizations should support and help in the creation of social ties between team members distributed globally and use necessary re-sources to ensure that the human aspects of collaboration are addressed \citep{Kotlarsky_2005}. Globally distributed teams might experience cultural and lingual issues as well as a lack of trust in the professional qualities of a remote colleague \citep{Asp_2021}. These issues can be addressed by more frequent interactions, like team meetings \citep{Battin_2001}. \cite{Ebert_2001} suggest that international teams and rotating management responsibilities across remote sites are other means to build common trust and understanding with colleagues.

In summary, this study addresses the issue of hindering factors in effective knowledge management and provides recommendations on how to mitigate them in the context of a singular case study organization. Despite the fact that there are several studies that list such factors \citep[e.g.,][]{Sharp_2003,Fischer_2001}, as well as tangential studies such as the listing of issues in KMSs \citep{Alavi_1999}, the results from a systematic secondary study \citep{Asrar_2016} indicate that the issues in effective KM are diverse and context-specific. That is, the hindering factors of KM identified in one context often fail to generalize to other contexts such as different industries, geographically distributed work, culture, and size of the organization. Arguably, this calls for more in-depth case studies to understand different contexts before general guidelines can begin to form. In the next section, we discuss the context of this particular study.

\section{Methodology}
\label{sec-methodology}

\subsection{Case description}
\label{sec-case-description}

The case company is a large international IT company with a history of transitioning from a hardware to a software company. The company has an open policy to modern and agile methods, and most of their teams are geographically distributed. In the summer of 2022, representatives from the company were contacted regarding a research project on knowledge management, with the goal of improving the company's knowledge management practices and culture between cross-functional teams distributed globally. The organization was chosen due to its size, the geographical distribution of teams, and willingness to participate in the research endeavor. 

To achieve this, a total of 50 employees with a range of experience, position in their team, and years in the company were interviewed in a total of 22 group interviews. The least number of years in the company was less than one year, whereas the greatest number of years in the company was over 25 years. Professional experience in the field ranged from several months to over 30 years and the positions in the team varied from managers to trainees. 38 participants identified as male and 12 as female, which is a reasonably fair illustration of the ratio between males and females working in the case company. We interviewed only peers in the same interview. The reason for interviewing very different employees in terms of years in the company, experience in the field, and position in the team was to avoid elite bias and gain the most accurate picture of the topic from each of the teams. The aim was to have three employees from each team in the interviews, but this was not always possible due to various reasons, and some of the interviews was only held with two interviewees. The approach was to contact the head of each department for their opinion on the most suitable candidates. No rewards were promised, and the motivation to participate in the study likely came from a genuine interest in the topic. 

\subsection{Data collection}
\label{sec-data-collection}

This study utilized semi-structured interviews with pre-prepared topics of interest, while also allowing for free discussion and clarification questions. The interview structure was designed based on a literature review, observations from the case company, and their needs (cf. \ref{sec-appendix}). The interviews were conducted using Google Meet software, and recorded, and transcribed using an in-house tool. Four main themes were discussed, covering the company's knowledge, knowledge needs, issues with knowledge management, and possible solutions. The interviews were conducted in September and October 2022, with an average duration of 48 minutes. Each participant was shown a privacy statement regarding how the data would be analysed, reported, and shared, and that participant anonymity would be preserved. Participation was based on informed consent, and was voluntary without any particular incentives besides helping the research endeavor. The interviews were conducted in a similar fashion to individual interviews, with the exception that the interviewer ensured that each of the participant was given the opportunity to express their opinions. The goal of the small group interviews was not to form a consensus within the group, but to provide a more efficient way of interviewing several participants when compared to individual interviews. 

\subsection{Data analysis}
\label{sec-data-analysis}

The interview data was automatically transcribed by a digital assistant. These transcriptions were then compared to the recordings that were recorded during the interviews. Some adjustments had to be made to the automatically produced transcriptions from the basis of interview recordings. The transcriptions were also cleaned based on a clean verbatim transcription style which allows removing non-speech sounds, repetition of the same word, false starts, etc. This makes the transcribed text easier and faster to read without compromising the precision or reliability of the transcriptions.

This transcribing process resulted in 114 pages of text. Because of the sensitive nature of the interviews and the topics discussed, the case company decided that no external tools will be used to avoid any conflicts in data security. Thus, per their request, the coding and analysis of the interview data was done manually, within the company's premises on their servers and network.

This research initially gained its motivation from perceived observations of practical challenges. We chose to utilize conventional content analysis \citep{Hsieh_2005}. Per this method, the categories and the names of the categories flow from the data instead of being thought of beforehand. This allows the researcher to have a less biased view of those categories and the names of the categories and is not limited or influenced by categories that have been provided by possible previous literature \citep{Hsieh_2005}.

\begin{figure}
  \centering
  \includegraphics[width=0.95\linewidth,keepaspectratio]{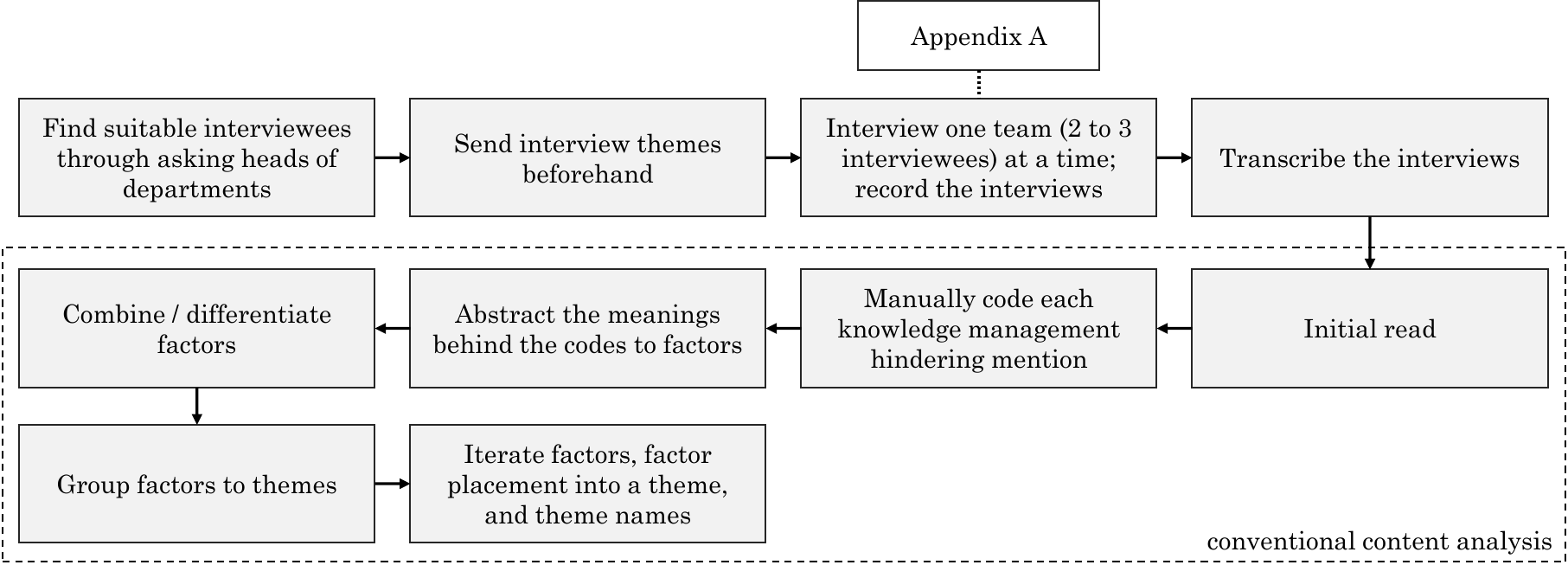}
  \caption{A summary of the data collection and analysis steps}
  \label{fig-process} 
\end{figure}

The coding process started with the first author reading through the transcripts to achieve a better overall understanding and to gain impressions. Then the first author carefully studied the transcripts iteratively as the coding cycles revealed an increasing number of initial codes. During this part of the process naming the codes would be in constant development since codes were merged and renamed based on the transcripts. Nearly 700 knowledge management issues were found in the transcripts that were then codified into 44 codes, as one of the objectives of the method is to formulate a higher-level understanding of the phenomenon studied. These 44 codes represented the hindering factors in knowledge management within and between teams in the context of software development (RQ1). These 44 codes were then grouped into 5 themes based on their estimated origin, and to give clarity to the results. Finally, these 44 codes were grouped into five categories from which the five recommendations were drawn (RQ2). The whole analysis was performed solely by the first author, and merely supervised by the second author. Fig.~\ref{fig-process} summarizes the data collection and analysis steps.

\section{Results}
\label{sec-results}

\subsection{Overview}
\label{sec-res-overview}

\begin{figure}
  \centering
  \includegraphics[width=0.35\linewidth,keepaspectratio]{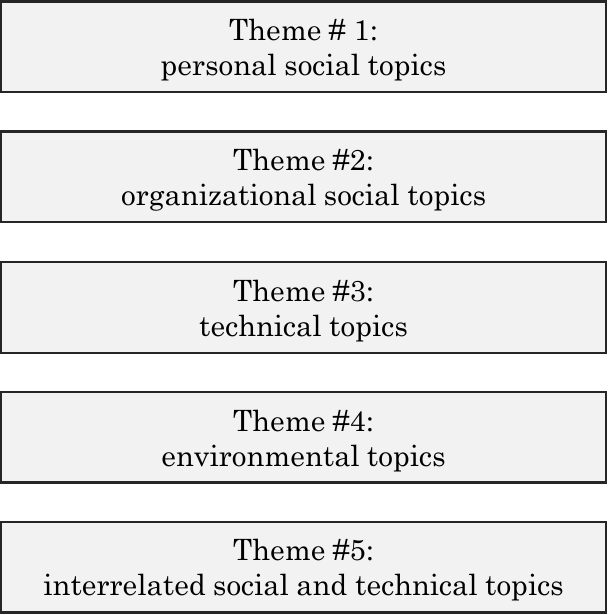}
  \caption{The five high-level themes of factors hindering effective knowledge management}
  \label{fig-hindering} 
\end{figure}

Fig.~\ref{fig-hindering} summarizes the findings on the themes hindering effective knowledge management. Some themes can have a negative impact on knowledge sharing and management in an organization. Other themes, although positive in nature, can also negatively affect knowledge sharing and management due to insufficient resources, lack of attention, or insufficient presence. In the case organization, some themes have been recognized, and measures have been taken to address them, but the results have not met the expectations. The next sections provide details and quotations about the themes.

\subsection{Theme \#1: Personal social topics}
\label{sec-res-personal}

\begin{figure}
  \centering
  \includegraphics[width=0.90\linewidth,keepaspectratio]{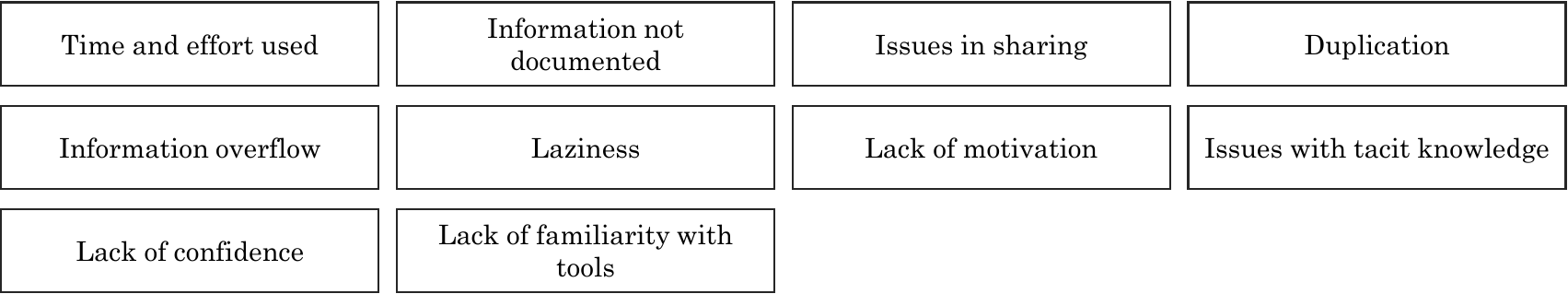}
  \caption{Factors hindering effective knowledge management under the theme \textit{personal social topics}}
  \label{fig-factors1} 
\end{figure}

The factors belonging to this theme are summarized in Fig.~\ref{fig-factors1}. A theme in this group was \textit{time and effort} used in relation to knowledge management. The majority of the interviewees expressed that they value their time highly and thus choose their tasks based on what is a good use of their time and is certain effort worth the time invested in it. It was recognized that sharing knowledge quite often means \textit{documenting} it somehow and then \textit{sharing} it with others. However, transferring knowledge into documents that others can find and understand was seen as burdensome because it requires unnecessary effort from both the owner and the receiver of knowledge.

\begin{quote}
\textbf{Time and effort}:
\textit
{
``Sometimes the data needs to be put in a document and this takes time and effort from both sides.''
}
\end{quote}

It was assumed before the interviews that documenting knowledge would be seen as `too much effort', but it was evident that also finding that knowledge was time consuming in many instances.

Some interviewees valued their time so highly that they did not see the purpose to participate in knowledge sharing if it was not a part of their job description.

\begin{quote}
\textbf{Issues in sharing}:
\textit
{
``Nobody wants to use their time on that [knowledge management] if that's not their job.''
}
\end{quote}

Another common theme mentioned in multiple interviews was the fact that lot of \textit{information is not documented} for number of reasons. Interviewees were often asked about tacit knowledge and how it is intended to be collected, transferred, stored, and retrieved. Common view to this is that tacit knowledge is not documented because it is hard or impossible to do so. However, this was not seen as the primary reason why tacit knowledge is not consistently documented in the case organization. Talking about sharing knowledge in general within the organization, it was highlighted that employees might not know what kind of information or knowledge they are expected to capture, store, and share.

\begin{quote}
\textbf{Information not documented}:
\textit
{
``Today I have to search in [different tools], wherever and it takes hours to find the correct information.''
}
\end{quote}

\begin{quote}
\textbf{Information not documented}:
\textit
{
``People need to understand, depending on their job, what information they are expected to capture and where they should feed the information.''
}
\end{quote}

Codifying process also identified issues with sharing when it comes to individuals. This is closely related to individual behavior where knowledge is kept from others for one reason or another. Most often this was not seen as intentional withholding of knowledge but rather as lack of motivation to be active about sharing knowledge. It was recognized that a lot of information is attainable and acquirable all the time, but the process of sharing that acquired knowledge is faulty. Couple of the teams felt like teams have developed a habit where knowledge is not openly shared without a reason. This was explained by the fact that it would be very wearying to share everything or to know all the people to share with.

\begin{quote}
\textbf{Issues in sharing}:
\textit
{
``This company [is] only sharing information when it's needed or requested.''
}
\end{quote}

\textit{Duplication of information} was also mentioned multiple times. Same information is stored to different locations and then possibly updated in different instances and times, which causes a lot of confusion. Finding the correct information seemed to be a major challenge for many.

\begin{quote}
\textbf{Duplication}:
\textit
{
``[The information is] somewhere and the challenge is to know where it is. We have several sources of information on the same topic.''
}
\end{quote}

Duplicating the information is due to a lack of knowledge on where to store which information, but also due to access issues and usability of tools. Different stakeholders use different tools for different purposes and want to have information relevant to them in the tools they use, which automatically causes duplication.

Sometimes the issue is that with all the accumulating information spread across different tools and colleagues' minds, there's an \textit{overflow of information}. There are so many sources where knowledge can be gained that it turns extremely difficult to know which source and which piece of information is really relevant to a single individual.

\begin{quote}
\textbf{Information overflow}:
\textit
{
``For me it's maybe too much information and finding what kind of information is valuable for me.''
}
\end{quote}

When it comes to people's intentions to act, personal motivation and willingness to play a role. The reason for contacting a colleague rather than looking the information from available databases was asked in several interviews. Most of the time the interviewees saw that asking a colleague is faster and more efficient, but one participant went a bit deeper with the question and thought that plain \textit{laziness} could be the underlying reason why colleagues get contacted more often than they maybe should.

\begin{quote}
\textbf{Laziness}:
\textit
{
``Sometimes it's about people really not knowing where to go and asking sincerely and sometimes people are just lazy and prefer contacting people for direct answers.''
}
\end{quote}

Improving knowledge management and sharing within the organization seemed to be a common view among interviewees, but some concerns were raised when it comes to motivating colleagues to actively participate in the change. Some issues were seen in the current way of communicating and willingness to support change and addressing these \textit{motivational issues} were seen as key objectives.

\begin{quote}
\textbf{Lack of motivation}:
\textit
{
``The will of changing things and the will of better communicating, not just through digital platforms, are prerequisites for me in order for the change to work in the future.''
}
\end{quote}

\begin{quote}
\textbf{Lack of motivation}:
\textit
{
``The question is how we could get the people to support this initiative of having better sharing and having common rules and having common ways of working.''
}
\end{quote}

The case organization has undergone some major changes in the past decade and that still affects the way of working and doing things. Some are still finding the older ways of working better whereas others prefer the new ways, which can create arguments and motivational issues between teams and colleagues. One participant noted how this is poor from a motivational point of view.

\begin{quote}
\textbf{Lack of motivation}:
\textit
{
``Especially from the motivational point of view, this is pure horror to cook those old things, old files rather than doing the things which are our future.''
}
\end{quote}

Regarding \textit{tacit knowledge}, the interviewees recognized the importance of such knowledge but felt like the perceived importance of tacit knowledge does not match the actual efforts to benefit from such knowledge.

\begin{quote}
\textbf{Issues with tacit knowledge}:
\textit
{
``In many cases [tacit knowledge] is more important than the official information. It's a big challenge to not miss this valuable information.''
}
\end{quote}

Lack of confidence was mentioned by a few individuals when asked about factors affecting knowledge sharing and management negatively. Sometimes this has to do with not knowing how to do something, but more often not knowing if something is allowed. Distributed software development is dependent on different teams and stakeholders working together, and very often some of the work is done outside of the organization by third parties like suppliers. Although collaboration is promoted, encouraged, and often necessary, some things have to keep within the organization for various reasons. Making a \textit{confident} decision on whether something can be shared at any given time or not seems to cause issues in knowledge sharing.

\begin{quote}
\textbf{Lack of confidence}:
\textit
{
``People are sometimes afraid that we share externally some information that we should not share because is not ready or because it's purely internal. People should also be more confident.''
}
\end{quote}

\begin{quote}
\textbf{Lack of confidence}:
\textit
{
``Sometimes I feel that people are a bit scared of doing things.''
}
\end{quote}

Related to confidence, familiarity with the tools in place was also identified in the coding process. As mentioned earlier, changes in the organization may introduce the introduction of new tools. \textit{Familiarizing on how to use any new tools} can be difficult with limited time on hand, and this is why even the tools implemented long time ago can cause challenges for employees who are not regularly using them. One interviewee admitted that even though some tools have been used for years, they are so complex that one does not have enough time to thoroughly learn to use those tools.

Employees not working at the organization's main sites used common tools so rarely that they tend to forget how to use them and instead come up with ways to collaborate that might be unofficial. This increases the amount of work colleagues at the main sites have to do when handling these unofficial requests and communication attempts.

\begin{quote}
\textbf{Lack of familiarity with tools}:
\textit
{
``People [in remote sites established as first contact points for customers at those locations] are not regularly using our tools, our databases, our information pages and those kinds of things and so they are basically forgetting how to use the things and they are often starting as a newcomer even when they have been in the company for years.''
}
\end{quote}

In summary, personal social topics represented the themes that are dependent on an individual's actions, capabilities, motivation, and feelings on knowledge sharing and management. Personal social topics were mentioned by every team across the organization when asked about factors that are or could be hindering knowledge sharing and efficient management within the organization.

\subsection{Theme \#2: organizational social topics}
\label{sec-res-organizational}

\begin{figure}
  \centering
  \includegraphics[width=0.90\linewidth,keepaspectratio]{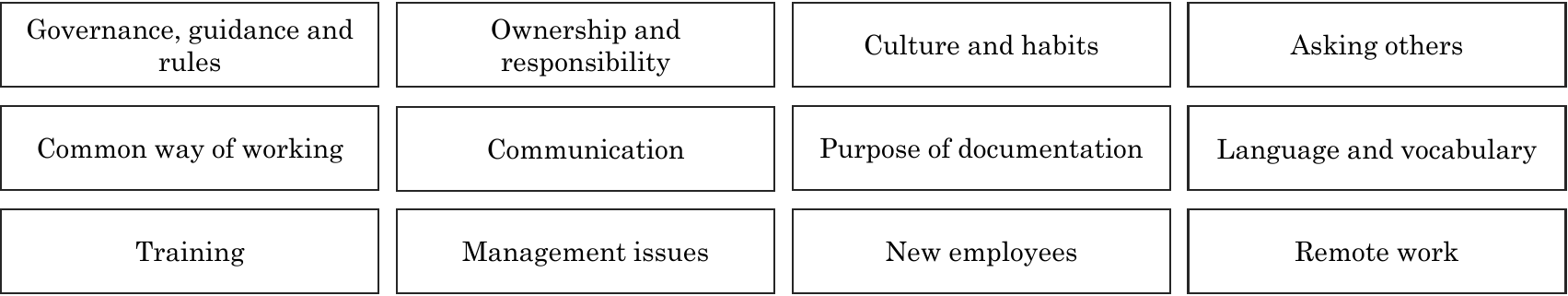}
  \caption{Factors hindering effective knowledge management under the theme \textit{organizational social topics}}
  \label{fig-factors2} 
\end{figure}

The factors belonging to this theme are summarized in Fig.~\ref{fig-factors2}. A commonly mentioned theme on organizational social topics was the \textit{lack of governance, guidance, and rules} related to knowledge management. The size of the organization dictates that rules need to be agreed on to enable successful business activities. This also applies to knowledge management. It was acknowledged that there are some guidelines and rules related to information management, but they were not commonly followed or even recognized by different team members. A need for overarching guidelines and governance to unite teams and team members with knowledge management was often asked for in the interviews.

\begin{quote}
\textbf{Governance, guidance and rules}:
\textit
{
``What we lack are some recommendations about how we should handle each type of data to have something homogeneous between teams and have similar ways of working and have some rules because today anyone is allowed to do whatever they want.''
}
\end{quote}

Changes in the business as well as in the technologies and tools have made it more difficult to set up binding rules. A constantly evolving landscape demands dynamic solutions that can be adapted to changes. With the rapid pace of changes, it has been challenging to keep rules and guidelines up to date throughout the organization and some of this lack of governance has been intentionally done to give more freedom and agency to employees to define new guidelines and rules. The challenge in this method is that everyone has come up with rules of their own and there has been little success in uniting these rules and guidelines over one governance.

\begin{quote}
\textbf{Governance, guidance and rules}:
\textit
{
``We really didn't have any training how to do so, and now people are just making up the rules as they go.''
}
\end{quote}

In some modern organizations, there are data owners or data custodians whose job is to look after the data possessed by the organization. In some organizations these responsibilities are embedded into job descriptions of people that deal with other responsibilities as well. Sometimes the \textit{responsibility for data, information, and knowledge} that an organization possesses is unclear and falls unto those that are willing to put in the extra effort to act as data custodians.

\begin{quote}
\textbf{Ownership and responsibility}:
\textit
{
``We are still struggling a little bit in making it clear that it's not the tool that dictates the ownership of data but really the teams.''
}
\end{quote}

Every organization has its own \textit{culture and habits} which can only be learned in practice. Organizational culture has a strong influence on knowledge sharing behavior as well as managing knowledge. Participants felt like some of the habits teams have developed are going against the nature of knowledge sharing. For example, knowledge is rarely freely shared, but only in cases it is seen as absolutely necessary.

\begin{quote}
\textbf{Culture and habits}:
\textit
{
``I'd say, it's not shared at all or in very few cases and very specific cases.''
}
\end{quote}

On the other hand, habits often sprout from a prevalent culture that has been molded through the years. The dynamic working environment and software development in general encourage changes and improvements and these influence the organizational culture. However, effects in organizational culture have been more ad hoc rather than systematic when it comes to intelligence culture. Intelligence culture promotes the capabilities and abilities to acquire, share, and benefit from possessed intelligence. A notable number of participants expressed their desire to promote an intelligence culture where reacting to market changes, competition's new features, and customer preferences would be rapid and knowledge-based. However, one participant speculated that maybe the value of intelligence culture is not yet fully recognized in the organization.

\begin{quote}
\textbf{Culture and habits}:
\textit
{
``At the end of the day, I'm pretty much convinced that this is primarily a cultural problem that we are lacking the intelligence culture. We are not used to do that kind of work, we don't maybe think that it's important, we haven't built that kind of work in the job descriptions and so on.''
}
\end{quote}

\textit{Asking others} for information and knowledge is common in knowledge-intensive fields. Understandably some knowledge is only available by asking the experts but asking your colleague for certain information can often turn into a habit. Constantly contacting someone instead of getting the information for example from a public database or company wiki page might take less time from the one asking, but it does take extra time from the one answering these requests and questions. Several participants acknowledged that too often colleagues are asked about things that are easily available and documented in a database or wiki page etc. The easiest explanation for this would be that asking others for knowledge saves time and effort and that's why one does so.

\begin{quote}
\textbf{Asking others}:
\textit
{
``That's probably because people know that I'm the key contact person for [a certain tool] so it's easy to get the answer quick.''
}
\end{quote}

However, when most employees prefer to contact colleagues directly when looking for information, it becomes increasingly difficult for newcomers and part-time workers to find the information they are looking for. When sharing knowledge is just between people, one is required to know enough people to find the answers needed, which would require a \textit{common way of working} to ensure everyone has the support they need for knowledge sharing.

\textit{Communication} plays a crucial role in sharing knowledge. A lot of situations re-quire instant action to inform appropriate stakeholders, when documenting the information to public sources is not enough. Last-minute changes, crises, tight schedules, or changes in the teams could be instances where information needs to be conveyed effectively to all appropriate stakeholders through direct communication channels. A lot of information was seen to be communicated unofficially which means that being out of the office, or being at the wrong coffee table or wrong part of the office could mean that one misses critical information.

Multiple participants raised a worry about internal communication. They stated that because of the spread-out nature and size of the company, many important pieces of information are not communicated officially internally and the only option to acquire that information is to be at the right discussions with the right people at the right time, which is in no measure effective or reasonable.

\begin{quote}
\textbf{Communication}:
\textit
{
``The way we get information and communicate is crucial and we are not yet doing it completely. There are many things to consider. There is the repository and the documentation that is produced in the context of the project. Then there is internal communication we need to do further.''
}
\end{quote}

Sharing knowledge through direct interactions can be easier than documenting it. When interacting with someone in person, it is possible to tailor the conversation to their specific needs. However, documented information can be accessed by colleagues and teams from different functions, who may not share the same level of understanding about the topic. This can make knowledge management challenging. One obstacle to effective knowledge management is understanding the \textit{purpose of documenting} each piece of information. It is important to consider who might need the document, what their level of knowledge is, and how best to explain the information. In software development, multiple functions with different areas of expertise work together, which makes it crucial to understand the purpose and audience of each document.

\begin{quote}
\textbf{Purpose of documentation}:
\textit
{
``And this is now the learning curve also for us, and we need to [...] understand that this is not only for us. This is not only, let's say, an R\&D document, it is also a document that is more widely available inside our company.''
}
\end{quote}

Many international companies have some kind of issues related to \textit{differences in language}. Internal vocabulary within the company was seen as a challenge. Even if the organization's official language is English, people tend to name products, projects, services, technologies differently – especially when using a language that is not their native language. Sometimes using the wrong keyword to search for information might be the reason why no information is found on that specific topic.

\begin{quote}
\textbf{Language and vocabulary}:
\textit
{
``The same thing can be named in several ways and when you're trying to search for something you have to imagine how someone else could have named it.''
}
\end{quote}

In knowledge management, it might be intuitive to focus on the one producing, sharing, and transferring knowledge. On the other side of knowledge management is the other person searching, receiving, and using the knowledge provided by someone else. Providing \textit{training} that considers both sides of knowledge management is something that the participants thought that was missing from all of the employees, especially the newcomers.

\begin{quote}
\textbf{Training}:
\textit
{
``I have not seen training about data and information management, data sharing, good practices, or guidelines. It goes both ways; it goes from the people that look for data as well as the people who make this data available.''
}
\end{quote}

Knowledge management is not something one individual or even a group of individuals can effectively promote in a large organization, but it requires a strong commitment from the \textit{management}. Lack of support from the management level often causes poor individual efforts to improve the situation. Few managers stated that the organization has all the needed capabilities to improve knowledge management but is missing a concrete plan on how to do so and the driving force from management.

\begin{quote}
\textbf{Management issues}:
\textit
{
``Actually, I don't see it difficult to gather that information but structuring and keeping it up to date all the time would be difficult. There could be a way to organize that information, but I don't have an idea of how this could be done in practice.''
}
\end{quote}

Some participants were worrying about the experience for \textit{newcomers}. Whenever a newcomer starts working, there is a large amount of information to be absorbed to handle daily tasks, and knowledge management might not be the first thing priority when trying to grasp new responsibilities. Few of the participants stated that providing an induction to newcomers where knowledge management-related issues would be addressed is necessary.

\begin{quote}
\textbf{New employees}:
\textit
{
``We are so busy with what we do that whenever newcomer comes in [...] we don't start the introduction in the organization by telling how to use the tools.''
}
\end{quote}

On top of working in globally distributed teams and organization, the COVID-19 pandemic caused a massive increase in \textit{remote working}. Similar to working in different countries, working remotely can cause issues in communication and knowledge sharing.

In summary, organizational social topics were the most visible in collaborative organizational settings where multiple individuals are involved. These themes were more affected by the actions and decisions of multiple individuals rather than one single individual as could be the case with personal social topics.

\subsection{Theme \#3: technical topics}
\label{sec-res-technical}

\begin{figure}
  \centering
  \includegraphics[width=0.90\linewidth,keepaspectratio]{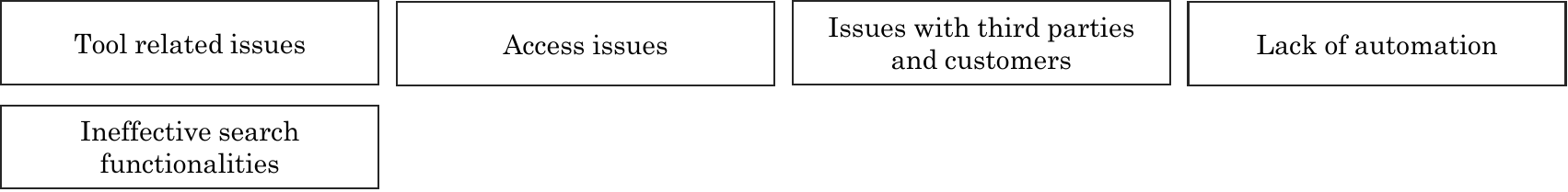}
  \caption{Factors hindering effective knowledge management under the theme \textit{technical topics}}
  \label{fig-factors3} 
\end{figure}

The factors belonging to this theme are summarized in Fig.~\ref{fig-factors3}. \textit{Tool related issues} were frequent among all the themes presented in Fig.~\ref{fig-hindering}. Even though the need for multiple tools was acknowledged widely among participants, many felt that having so many tools distracts knowledge sharing.

\begin{quote}
\textbf{Tool related issues}:
\textit
{
``We have so many tools and as said you don't know which tool the team is using. There's no way of using any kind of search capabilities because you don't know the tool.''
}
\end{quote}

Others argued that correct tools are in place for successful performance, but the issue is the synchronization between the tools. Having important information in several tools that do not communicate together requires lot of manual work.

\begin{quote}
\textbf{Tool related issues}:
\textit
{
``I think we have all relevant components but they are not in sync. We are missing a lot of data due to that. The information is really scattered everywhere. We cannot base our decisions on historical data and there's absolutely no sync between the tools.''
}
\end{quote}

However, despite having multiple tools for knowledge management, they do not always fit a certain purpose or task. Thus, existing tools need to be used creatively and to the best of their capabilities. Sometimes this causes issues because using tools for what they are not meant to be used will eventually result in problems like poor structure of information or validity of information.

Several factors might restrict \textit{access} that individual employee has to different tools. When asked about accessing relevant information the interviewer was reminded that all data cannot be accessible to everyone. Some information is classified as confidential or to be seen only by certain teams or project members which causes restrictions on access rights. Other access restrictions are due to budgetary reasons; every license to access certain tools costs money and thus access is only given to those who have an absolute need to access information in that tool regularly. Sometimes getting access simply takes some time or access rights are not requested immediately for newcomers. It is also possible that the creator of documents restricts the availability of them either on purpose or by accident. One participant noted that especially as a newcomer it seems to be more difficult to access documents created by someone else than some general company documents.

\begin{quote}
\textbf{Access issues}:
\textit
{
``I think there's an assumption in your question that is not validated yet and it's that all the data is accessible by everyone for any kind of purpose, and this is not the case.''
}
\end{quote}

\begin{quote}
\textbf{Access issues}:
\textit
{
``Something that I have experienced during the past few months is that I don't have access to certain tools or even to certain [internal wiki] pages.''
}
\end{quote}

Organizations often work with \textit{customers, suppliers, and other third parties} regularly and need effective ways to collaborate and cooperate with those stakeholders. Having a robust system to collaborate and communicate with external stakeholders can be a challenge because all parties have their demands and requests. Sharing information with these parties can be difficult, especially if the shared information is classified or for certain eyes only.

\textit{Automation} in knowledge management is not simple or easy, but the lack of automation forces organizations to invest in people who manually provide reports and analysis for decision-makers, which takes time and money. Creating these supportive materials for decision-making is no longer a question of `should we' but a question of `how'.

\begin{quote}
\textbf{Lack of automation}:
\textit
{
``Everything now is based on us asking and being active. It's very much like manual or human work we need to do.''
}
\end{quote}

The quest of \textit{finding information} from available sources and tools requires the use of the search functionalities provided. The issue with these functionalities is that finding specific information can be challenging for several reasons. Using the wrong wording, searching from the wrong tool, using the functionality wrong, not using search filters, or searching for information that is restricted can all be reasons why finding the correct information can seem to be impossible. Participants gave a few examples of this frustration. If one seeks information from an internal tool that is restricted, it will not even turn up in the search, leaving the person to think that the document does not exist when it's only restricted for the time being. Another example is that one is not sure for which tool information should be searched and because there's no search engine that would search from all the tools, one must manually go through all the possible options. Wrong wording was briefly discussed earlier, but even with the correct keywords one might find tens of thousands of results only to be confused about the correct in-formation.

\begin{quote}
\textbf{Ineffective search functionalities}:
\textit
{
``You'll have 16,000 search results, but you don't find what you need because there's no answer related to the thing you are looking for.''
}
\end{quote}

In summary, globally distributed software development requires extensive knowledge sharing and effective communication that has to be completed through different tools because face to face interactions is limited. This dependency and importance of tools is reflected in the number of concerns directed to tool related issues. First, large organization has a vast variety of tasks, teams, and responsibilities, which require a variety of tools to be in place.

\subsection{Theme \#4: environmental topics}
\label{sec-res-environmental}

\begin{figure}
  \centering
  \includegraphics[width=0.90\linewidth,keepaspectratio]{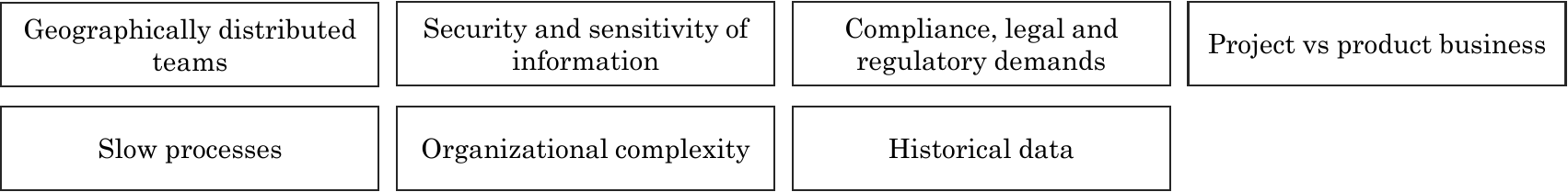}
  \caption{Factors hindering effective knowledge management under the theme \textit{environmental topics}}
  \label{fig-factors4} 
\end{figure}

The factors belonging to this theme are summarized in Fig.~\ref{fig-factors4}. In the case organization, \textit{distributed teams and regions around the world} were working together to plan, produce, sell, and maintain software products and services. They communicate and collaborate daily in different ways and settings. Based on the interviews, the biggest challenge with working remotely with colleagues is the fact that casual information sharing and brainstorming during coffee breaks and hallway conversations are almost non-existent.

\begin{quote}
\textbf{Geographically distributed teams}:
\textit
{
``I think this is the throwback of working remotely, these casual conversations and brainstorming does not exist.''
}
\end{quote}

When a team or a site has to work remotely with others, it is easier to come up with own ways of working and stop sharing knowledge that feel small or insignificant. This creates a snowball effect where these teams and sites develop a self-sufficient way of working where others are not necessarily needed. This directly affects knowledge sharing and cross-team collaboration.

\begin{quote}
\textbf{Geographically distributed teams}:
\textit
{
``Those are somehow by default quite self-sufficient, and this is a bit of creating a set-up where we have many different islands [...] kinds of teams that are at least a little bit in isolation from other teams.''
}
\end{quote}

Working with different customers and stakeholders sometimes causes restrictions on the information that is passed around. The \textit{security and sensitivity} of documents and information need to be taken seriously from both sides of the interaction. In particular cases, access to certain information needs to be restricted, information transfer needs to be encrypted, and communication needs to be secured. These restrictions cause additional work, but also pose requirements to the tools being used and the way of working.

\begin{quote}
\textbf{Security and sensitivity of information}:
\textit
{
``We have a lot of sensitive information that we cannot put everywhere [...] this is very complex.''
}
\end{quote}

\begin{quote}
\textbf{Security and sensitivity of information}:
\textit
{
``Then of course, there are some sensitive data, also, nationally sensitive or atomized information, which is secret and thus handled in a different way. There, of course, the information is much more restricted.''
}
\end{quote}

Related to the sensitivity of information, there are \textit{legal and regulatory demands}, like compliance rules, that need to be obeyed in everything – including knowledge management. For example, some product information cannot be shared across country borders, or certain customers want to limit access to a minimum when it comes to solution integration or implementation to their current systems. Also, sharing confidential information about competitors or their products and services, even if attained by accident from competitors, can be seen as corporate espionage and thus break compliance rules. Breaking these rules or legal regulations can cause heavy punishments for the organization. These cause hindering factors to knowledge sharing and management and might motivate individuals to keep information to themselves rather than sharing it, because they are not sure if it can be shared or not.

\begin{quote}
\textbf{Compliance, legal and regulatory demands}:
\textit
{
``We have very limited access to only certain people, in order not to breach the confidentiality obligations.''
}
\end{quote}

\begin{quote}
\textbf{Compliance, legal and regulatory demands}:
\textit
{
``As we know there are sometimes regulatory and legal reasons why some things just can't be shared.''
}
\end{quote}

Software projects have been common in the past years where software is designed to fit set requirements from the customer and then modified and maintained according to customer preferences. Nowadays, however, the \textit{project-based approach} is challenged by the product-based approach. The product-based approach means that software development is focused on developing the product according to the markets, not just individual customers. This is a more permanent way of thinking about products and services because software is not developed to match clients' needs but to match the market domain’s needs. This enables a one-fit-for-all approach and changes are initiated from market and domain changes rather than the client's request. However, the case organization is still in the middle of a transformation coming from a project-based organization to a product-based organization, which causes issues with knowledge management and sharing. Clarifying the approach was seen as essential for the future success of the organization.

Technology, markets, and software development is changing fast, and organizations are doing their best to stay up to date. A larger organization often means \textit{slower processes}. Participants agreed that many things could be more up-to-date, but at the same time, several participants stated that it is nearly impossible to stay up-to-date in all domains. Some even claimed that staying up-to-date in all domains is not necessary but couldn't really define the domains where being updated would be most important.

\textit{The complexity of the organization} was also seen as a discouraging factor in knowledge management. This view would be understandable if this idea comes only from newcomers who haven't had enough time to get acquainted with the organization and teams. However, a participant who has been working in the organization for years said that one of the issues with knowledge management is the fact that complex organization causes confusion on who has what information and what should be shared with whom.

\begin{quote}
\textbf{Organizational complexity}:
\textit
{
``So, the setup is quite complex. And this is making this data management a complex exercise because there's no clear consolidation point and not always clear roles who is providing the needed information.''
}
\end{quote}
 
The case organization has been operational for a considerable amount of time and has collected data and information during the years. One of the issues for knowledge management is the question on what to do with the \textit{historical data}. It is hard to define how long documents should be updated or maintained, when they should be moved into archives, and where they should be stored in case they are needed somewhere in the future. Few of the participants noted that historical data cannot be just deleted but needs to be stored somewhere. However, historical data would have to be separated from the data that is currently up to date and does not pop up so easily in searches. This issue is still under process and causes issues in decision-making.

In summary, the organizational environment is something quite static and challenging to change. These topics related to organizational environment share similar attributes where they are relatively static themes that are challenging to change. The nature of these topics means that changing these situations directly is likely not an option but rather having workarounds to ease knowledge management.

\subsection{Theme \#5: interrelated social and technical topics}
\label{sec-res-interrelated}

\begin{figure}
  \centering
  \includegraphics[width=0.90\linewidth,keepaspectratio]{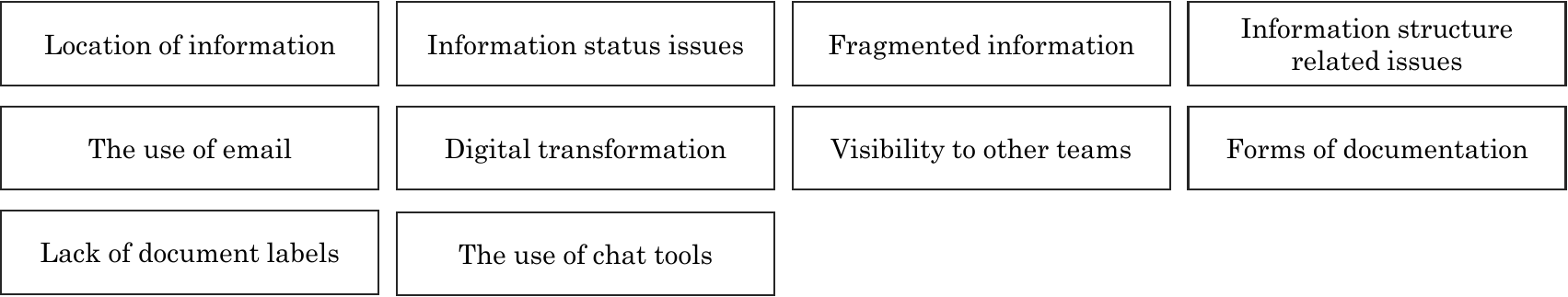}
  \caption{Factors hindering effective knowledge management under the theme \textit{interrelated social and technical topics}}
  \label{fig-factors5} 
\end{figure}

The factors belonging to this theme are summarized in Fig.~\ref{fig-factors5}. One of the major issues with knowing the \textit{location of the information} is effective informing. One participant expressed frustration on how information is stored \textit{somewhere}. New employees store information in new places, new tools are being introduced with new needs and procedures, and different colleagues store information about products and services in different locations. These locations are not effectively communicated, so it is up to the seeker to find information.

\begin{quote}
\textbf{Location of information}:
\textit
{
``The problem with this is that we have [internal wiki] pages that are not communicated, the links are not being shared.''
}
\end{quote}

Almost all of the participants answered that \textit{information status} is a topic that crosses their minds regularly when dealing with knowledge and information. Participants had difficulties in identifying which information is the most recent and up-to-date which forces them to track down a colleague that might know something about it.

\begin{quote}
\textbf{Information status issues}:
\textit
{
``Right now, this information is spread in different places and it's extremely difficult to know where the information is, where the accurate information is, and be sure that it's up to date. Typically, we can find something in a collaboration tool. It's fine but we don't know if it's the last information if it's the reference information, if it's working information... sometimes we are totally lost.''
}
\end{quote}

\begin{quote}
\textbf{Information status issues}:
\textit
{
``It is very hard to understand is this really the latest version is this really the latest place and there's no kind of or not easy way how to recapture and basically find out.
''
}
\end{quote}

This also has a lot to do with the choices practitioners are making with their knowledge and possessed information. Some see documentation as a waste of time, because it becomes historical data instantly, whereas some others will stick to old information out of a habit or because they don't know where to find the latest information or care to request that information when it's needed.

\begin{quote}
\textbf{Information status issues}:
\textit
{
``I don't believe in documents because they are outdated the moment they are written.''
}
\end{quote}

\begin{quote}
\textbf{Information status issues}:
\textit
{
``Okay, what is the latest document? I have one stored on my hard disk. I'm using that one. I know it's from 2018, but is this really the latest one? I don't know. If you ask me where to find the newest template for NDA, I don't know. Who has stored it? Is it somewhere? Yes. But I don't know where to search from.''
}
\end{quote}

The possibility to store similar information to multiple different tools and the ability to choose freely which tool to use and when has results in scattered information. When asked about the reason, introduction of new tools and a lack of global overarching vision seemed to be the main causes of \textit{fragmented information}.

When knowledge is shared in person, it can be considered quite informal. When knowledge is conveyed through a document that can be read at any given time, it requires some kind of format. When asked about hindering factors related to knowledge management, \textit{information structure} came up multiple times. Participants explained that documentation must have some kind of structure so that it would be understandable to the reader and the information would be compatible with other sources. Having no structure means that everyone stores and shares information how they like and that causes difficulties. One of the potential reasons for not having a solid structure for storing and sharing information is the fact that information rarely stays static.

\begin{quote}
\textbf{Information structure related issues}:
\textit
{
``I agree that information exists, but it's gathered and presented in different ways. For this reason, it's challenging to compare the info because it's not stored/structured in a similar way. Not sure if the information is compatible because different sources present the facts differently.''
}
\end{quote}

An \textit{email} has proven effective for communication regardless of time and location. A lot of information is shared via emails daily as it is an effective tool to communicate both internally and externally. Despite the undeniable benefits, many participants mentioned email hindering knowledge management. One participant described how emailing can be used for the wrong purposes, which would later cause issues in finding information.

\begin{quote}
\textbf{The use of email}:
\textit
{
``I think the email is quite useful for technical discussions of things like that. But when  they are used for decision making while decisions should be made on more as a stable environment.''
}
\end{quote}

When asked for a reason why email is sometimes used too much and on the wrong things, answers were quite similar. Emailing is a faster option than other alternatives or there is no other alternative.

\begin{quote}
\textbf{The use of email}:
\textit
{
``Sending emails takes a few seconds whereas putting files into [a shared location] and asking the customer to download them from there takes a lot more time.''
}
\end{quote}

\begin{quote}
\textbf{The use of email}:
\textit
{
``It's easy to ask it through the email or in the chat [...] but for example [...] the messages are stored there for a couple of months so not forever. Then it's lost. When looking at the history of what has been done and what's the current status and then we are trying to see the information from the chat and it's not visible to everyone, it might not be found anymore or in the emails and that's a problem.''
}
\end{quote}

\textit{Digital transformation} in recent decades has changed the way technology is used, the way work is done, and the purpose of employees. Understandably these changes also affect knowledge management on many levels. New tools to be used and new ways of working to be familiarized is something that participants felt like would affect the way knowledge management is both perceived and promoted because it gives an opportunity to get rid of the old habits. 

Another factor that discourages employees to share knowledge with their colleagues is poor \textit{visibility to other teams}. Teams around software development need to work together tightly and be able to collaborate. This can become challenging with remote teams if there is no clear visibility between the teams and functions.

Some participants expressed that documentation in part is useless. They were worried if certain reports and documents are even read by anyone even though they have to be produced for standardization purposes. Having the right \textit{form of documentation} can be useful in instances like these and promote better knowledge management in a long run.

\begin{quote}
\textbf{Forms of documentation}:
\textit
{
``I'd have it properly written down because of this broken telephone tendency. There are a lot of people who have the need to have written documentation of the information that they need.''
}
\end{quote}

A few people found current \textit{labeling} tendencies within the organization poor and thus also discouraging from a knowledge-sharing point of view. Labeling or tagging documents and information would help in narrowing down searches, creating a structure for documents, and classifying the importance of documents easily.

In recent years, alongside email, \textit{chat} services have gained significant popularity. Chat functionality serves as a valuable tool for swift messaging and communication. However, akin to the concerns raised regarding email usage, some participants expressed apprehensions about the potential misuse of chat platforms and the consequent loss of information. Retrieving past discussions in chat settings poses challenges due to inherent difficulties in search capabilities, and messages may be subject to deletion after a certain time period. The presence of multiple chat tools further compounds the issue, as important information is frequently lost within various chat platforms, resulting in wasted time when attempting to locate past conversations that later prove essential.

In summary, interrelated social and technical topics illustrated themes where both technology and humans have a distinct effect on the topic. The location of information is dependent on both the technical capabilities of tools and human action. The technical side might cause restrictions on what kind of information can be stored where and when, whereas practitioners can decide which tool to use when storing and sharing their knowledge. 

\section{Discussion}
\label{sec-discussion}

\subsection{Theoretical implications}
\label{sec-disc-theoretical}

The findings of this study suggest that the importance of knowledge management is recognized widely across functions, but that recognition rarely turns into actual actions to improve knowledge management. It was discovered that personal social topics, organizational social topics, technical topics, environmental topics, and interrelated technical and social topics are hindering knowledge management improvement initiatives and actions. Solving these concerns would be crucial for the case organization to enable efficient improvements related to knowledge management and increase employees' motivation to participate and promote actions to improve organizational knowledge management.

The findings of this study indicate that managers and practitioners alike recognize the importance of knowledge management, knowledge sharing, and knowledge itself. This supports multiple studies conducted during the years \citep{Aurum_2008,Rus_2002,Ryan_2009,Ryan_2013}. Even though the participants admitted that knowledge management and sharing are important for them and for the organization, they also agreed that this perceived importance is not sufficiently reflected in the organization nor in the everyday activities of an employee. Despite considering knowledge sharing and management important, participants admitted that sharing, finding, and managing knowledge rarely match their perception of the importance of the matter. This aligns with the study conducted by \cite{Aurum_2008} except their statement was focused on the organization's passivity whereas this study focuses more on the individual's passivity in knowledge sharing and management.

Prior literature suggests that globalization in the IT sector has enabled offshore arrangements and partnerships with countries where the cost of labor is lower, which could in turn make employees withhold important knowledge on purpose as a way to ensure their employment \citep{Ebert_2001,Herbsleb_2001b}. Results of this study do not support this claim, as participants indicated their sincere willingness to share knowledge with their colleagues and did not express worry about losing their position or employment.

As noted earlier, knowledge plays a crucial part in the success of any IT company and is very highly valued. Value brings along responsibility which was indicated to be one of the main factors why knowledge sharing, and knowledge management are not as efficient as they could be. \cite{Wegner_1987} noted that when an organization does not have clear responsibilities, the chance of losing important information is significantly higher. The results from this study would seem to support this idea.

Multiple studies have shown that keeping a wide personal network, preferring social interactions, and locating colleagues with possessed knowledge is an effective ways to share tacit knowledge \citep{Aurum_2008,Bock_2005,Ryan_2009}, and thus managers should support the development of those relationships between colleagues and teams. However, participants indicated that sometimes this kind of approach to knowledge sharing and acquiring could be causing issues in a globally distributed setting. Having colleagues around the world means that personal social interactions are rare in person and mostly happen via phone calls or video meetings which don't compare to in-person meetings. Sometimes a colleague is not available for a call or meeting when information is needed as soon as possible, and this can happen often in software development projects. It would be important to have other means to gain information when that is not dependent on one individual or even a team. Only resorting to personal connections can discourage sharing and acquiring information to and from digital tools made for that exact purpose. Some participants admitted that asking their network for information is usually their go-to strategy and that is why they are not that familiar with the tools that store the very same information they are looking for. A better balance between these two sides was hoped for in the interviews.

The size of the organization can hamper any changes planned for knowledge management. Research suggests that as an organization grows, it develops formal systems, structures, and norms that slow its capacity to recognize changing conditions and react to them efficiently \citep{Leiblein_2009}. This study supports the idea of losing the ability to make quick adjustments to respond to shifting environmental factors when an organization grows. Participants admitted that the size and complexity of the organization make it harder to perform changes that require a lot of time and attention. A way to succeed in an effort to change the way knowledge management is handled is to have that initiative led by top management which can provide necessary resources. Although the size of the company can break down interpersonal relationships and communication \citep{Serenko_2007,Fores_2016}, creating a sustainable structure to promote knowledge management is a possibility when it is based on the organization's values and norms.

\subsection{Practical implications}
\label{sec-disc-practical}

\begin{figure}
  \centering
  \includegraphics[width=1\linewidth,keepaspectratio]{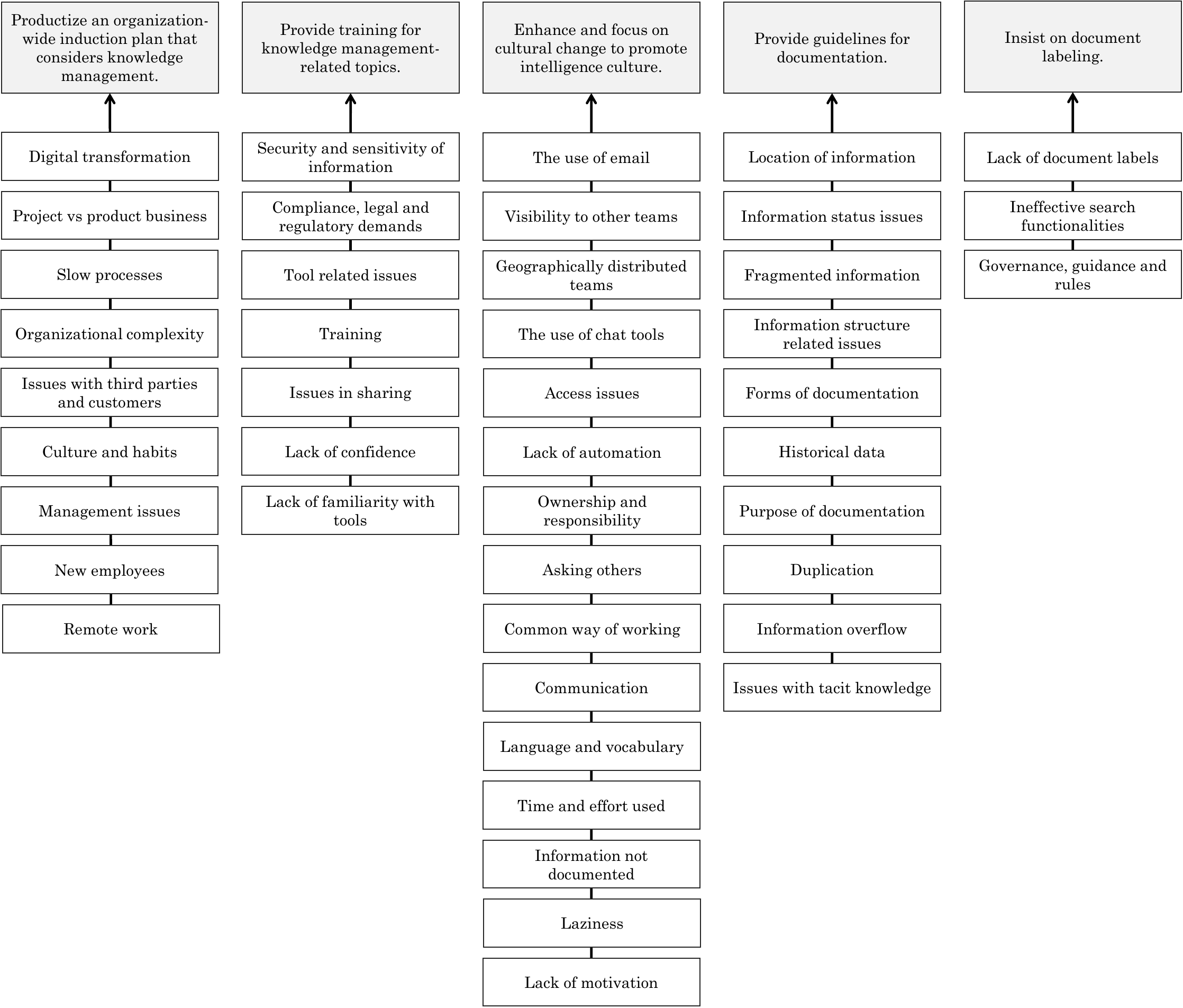}
  \caption{Five recommendations (top) drawn from the 44 knowledge management hindering factors identified}
  \label{fig-recs} 
\end{figure}

We provide five recommendations drawn from the interviews. These recommendations support the ongoing initiatives improving knowledge management in the organization by covering the main hindering factors not yet covered. These recommendations aim at transforming employees' interest and perceived feelings of importance towards knowledge into concrete actions. Fig.~\ref{fig-recs} summarizes the recommendations and illustrates from which hindering factors the recommendations were inferred from. Despite the structure of the figure, there is a level of overlap between the factors, and despite that the figure infers that one factor only contributes to one recommendation, the abstracted nature of the factors and recommendations effectively means that the connections are not as clear-cut. The recommendations are shortly discussed below.

Participants agreed that there is little material on knowledge management available within the organization and finding a way to access and acquire necessary knowledge can be overwhelming, especially for newcomers. One participant had started in the organization a couple of months ago and was still wondering where to find certain key information and how to use some of the tools that possess documentation. Instead of having an introduction to the topic, this new employee had to ask around, email unknown colleagues, and spend hours with different tools containing some knowledge. An organization-wide induction plan would not only give the opportunity to inform newcomers about knowledge management practices and culture, but it would also unify the way the employees work with knowledge and the tools dedicated to supporting knowledge management. The details of such a plan are highly context-dependent. Findings along similar lines have been presented in, e.g., \cite{Singh_2021} and \cite{Nakash_2023}.

Newcomers are not the only employees who struggle with sharing and acquiring knowledge on a daily basis. More experienced employees have old habits that might complicate knowledge management and wider networks within the organization which could mean that they manage and share knowledge only within their social circle or personal network. Participants felt like the reason for this kind of behavior most likely is the fact that no training has been provided on knowledge management topics. When there are no guidelines, resourceful people come up with their own way of doing things. When multiple employees around the globe come up with their own way of documenting, sharing, and acquiring knowledge the outcome can be complex. Providing training on how knowledge should be documented, shared, acquired, and maintained would help in simplifying knowledge management within the organization. Training should consider not only the donor and owner of knowledge but the receiver and searcher of knowledge as well. The training could instruct how to use available tools for knowledge management, how knowledge should be documented, how different sources of information could be bookmarked or flagged, and how information should be searched. This kind of training would have similar benefits to the induction plan but would be directed to all employees. \cite{Abubakar_2019} have presented similar findings.

An organization's culture strongly defines the way work is done and what is considered important. Participants expressed that the current culture does not support knowledge sharing as well as it could. It was indicated that intelligence culture, in which knowledge and benefiting from knowledge are in the middle of, was still lacking. Some participants suspected that intelligence culture is not seen as very important yet because it has not been actively promoted, it has not been included in job descriptions, and employees are not used to doing that kind of work. It was claimed that most of the employees don't fully understand the importance of knowledge and intelligence and that is reflected in the current situation and culture. Basing the culture around knowledge would automatize some knowledge-sharing, acquiring, and management practices. The successful cultural change would affect employees' motivation and capabilities to draw more value from available knowledge. Similar findings have also been reported previously \citep{Butt_2019,Kalkan_2008}. 

This study showed that there were several issues with knowledge sharing when it comes to documenting knowledge. Participants acknowledged how difficult it was to know which document contains valid information, where certain documents can be found, and if certain documents can be compared because they are presented in various ways. Multiple locations of information, spread out knowledge, document life-cycle management, and access issues were all mentioned by participants when asked about issues with knowledge sharing. The overarching issue seemed to be the lack of common guidelines that would guide the donor of knowledge as well as the receiver of knowledge. Guidelines for documentation could consider how often certain documents need to be checked and updated, what is the appropriate location to store certain documents, who is entitled to have access to certain documents, and how knowledge should be presented and structured. These guidelines could be included in training materials and induction plans as they would form the base for the other recommendations discussed. These findings support the insights provided by \cite{Almeida_2014}.

Trying to find a certain document out of the tens of thousands stored in the database might seem like an impossible task sometimes. Participants recalled how they were searching for a certain document with the correct keyword and the search engine offered over 16,000 search results. This complexity takes unnecessary time and effort from employees, which causes many employees to abandon any technical tools and prefer contacting their colleagues directly, taking their time on the matter. A participant recalled how a colleague could not find information from the internal wiki page and contacted this participant. Then this participant had to use their time to go to that very same internal wiki page to find that information so that they could provide the link to the information. A potential solution is to use labeling in the documents. Attaching an agreed label to each document helps in categorizing documents and filtering down search results. If labeling would be supported and insisted on documents could be searched with the keyword and label, thus potentially narrowing the search results from tens of thousands to a handful. These findings are in line with the results presented by \cite{Gunnlaugsdottir_2003} and \cite{Razzaq_2019}.

\subsection{Limitations and threats to validity}
\label{sec-disc-limitations}

\textit{External validity}: the selected research method causes some limitations to this study. The purpose of an interpretive study is not to generalize the findings but rather to explain individual phenomena. In this case, this phenomenon was studied in a single case organization, thus possibly limiting some of the results to plausible only in certain settings or organization. Even though interviewees were from several different continents and countries, they are all more or less accustomed to the culture and habits prevailing in the case organization. The majority of invited employees agreed to participate in the interviews. Some, however, refused and this could be considered a possible limitation. Those who agreed to be interviewed were voluntarily doing so and this could be seen as elite bias where only those employees who are active in knowledge management activities would participate.

\textit{Internal validity}: data were collected through semi-structured interviews where participants were free to express their opinions and thoughts. Given the sensitive nature of the topic being discussed, it is possible that some issues could have remained hidden because of interviewees' choice to withhold information.

\textit{Construct validity}: the data collection method could also act as a limitation because the researcher interpreting the interviews is also prone to misunderstand or misinterpret. Interviews were conducted by the same person who performed the transcriptions and interpreted the results. Interviews were transcribed manually per the case organization's request which leaves room for human error. The coding process was also done by the same individual manually. Manual work predisposes to more errors and a single researcher could have researcher bias.

\textit{Conclusion validity}: again, the data analysis by one researcher poses a threat to validity. It is possible that the researcher misinterpreted the narrative or thoughts of the interviewees, selectively reported some of the findings, and reporting incomplete or inaccurate findings. Qualitative research focuses on understanding context-specific phenomena, and generalizability of findings to other contexts or populations may be limited. Unfortunately, due to the case organization's data policies, it was not possible for the second author to analyze the data.

\section{Conclusion and next steps}
\label{sec-conclusion}

The objective of this study was to explore factors hindering knowledge management and the reasons why the importance of knowledge management is not reflected in practice. The results indicate several factors that hinder knowledge sharing and management, namely personal social topics, organizational social topics, technical topics, environmental topics, and interrelated social and technical topics. Although divided into groups, respective hindering factors can be connected and dependent on each other in different situations where one factor can cause another. The primary contribution of this study is in the identification and descriptions of hindering factors that discourage and prevent effective knowledge sharing and management within and between globally distributed software development teams. Recognizing, identifying, and understanding these factors is a prerequisite for addressing them properly and maximizing the potential to increase the quality and quantity of sharing and managing knowledge. The results indicate that the factors that hinder the application of knowledge management practices are lack of motivation, lack of clear job descriptions, and lack of familiarity and confidence in practical aspects of the organization and projects. Additionally, we provided five high-level recommendations on how to mitigate these factors that hinder effective knowledge management. Effectively, we recommended facilitating holistic knowledge management practices, training, promoting intelligence culture, and guidelines for documentation and document labeling.

Practitioners can benefit from understanding potential reasons for poor knowledge management, and researchers can benefit from understanding practitioners' issues caused by these identified factors to focus the direction of future research. In the future, the scientific field of knowledge management requires multiple case studies in order to understand the nuances of different knowledge management contexts. Furthermore, there exists a research gap of understanding which contextual factors such as culture, organization size and industry cause which positive and negative effects on effective knowledge management. 

\section*{Declaration of interest statement}
This research did not receive any specific grant from funding agencies in the public, commercial, or not-for-profit sectors.



\bibliography{mybibfile}

\begin{thebibliography}{87}
\expandafter\ifx\csname natexlab\endcsname\relax\def\natexlab#1{#1}\fi
\providecommand{\url}[1]{\texttt{#1}}
\providecommand{\href}[2]{#2}
\providecommand{\path}[1]{#1}
\providecommand{\DOIprefix}{doi:}
\providecommand{\ArXivprefix}{arXiv:}
\providecommand{\URLprefix}{URL: }
\providecommand{\Pubmedprefix}{pmid:}
\providecommand{\doi}[1]{\href{http://dx.doi.org/#1}{\path{#1}}}
\providecommand{\Pubmed}[1]{\href{pmid:#1}{\path{#1}}}
\providecommand{\bibinfo}[2]{#2}
\ifx\xfnm\undefined \def\xfnm[#1]{\unskip,\space#1}\fi
\bibitem[{Abili et~al.(2011)Abili, Thani, Mokhtarian and Rashidi}]{Abili_2011}
\bibinfo{author}{Abili\xfnm[ K.]}, \bibinfo{author}{Thani\xfnm[ F.N.]},
  \bibinfo{author}{Mokhtarian\xfnm[ F.]}, \bibinfo{author}{Rashidi\xfnm[
  M.M.]}.
\newblock \bibinfo{title}{The role of effective factors on organizational
  knowledge sharing}.
\newblock \bibinfo{journal}{Procedia Social and Behavioral Sciences}
  \bibinfo{year}{2011};\bibinfo{volume}{29}:\bibinfo{pages}{1701--1706}.
\newblock \DOIprefix\doi{10.1016/j.sbspro.2011.11.415}.
\bibitem[{Abubakar et~al.(2019)Abubakar, Elrehail, Alatailat and
  Elçi}]{Abubakar_2019}
\bibinfo{author}{Abubakar\xfnm[ A.M.]}, \bibinfo{author}{Elrehail\xfnm[ H.]},
  \bibinfo{author}{Alatailat\xfnm[ M.A.]}, \bibinfo{author}{Elçi\xfnm[ A.]}.
\newblock \bibinfo{title}{Knowledge management, decision-making style and
  organizational performance}.
\newblock \bibinfo{journal}{Journal of Innovation \& Knowledge}
  \bibinfo{year}{2019};\bibinfo{volume}{4}(\bibinfo{number}{2}):\bibinfo{pages}{104--114}.
\newblock \URLprefix
  \url{https://www.sciencedirect.com/science/article/pii/S2444569X17300562}.
  \DOIprefix\doi{https://doi.org/10.1016/j.jik.2017.07.003}.
\bibitem[{Agerfalk et~al.(2005)Agerfalk, Fitzgerald, Holmstrom~Olsson, Lings,
  Lundell and O~Conchuir}]{Agerfalk_2005}
\bibinfo{author}{Agerfalk\xfnm[ P.]}, \bibinfo{author}{Fitzgerald\xfnm[ B.]},
  \bibinfo{author}{Holmstrom~Olsson\xfnm[ H.]}, \bibinfo{author}{Lings\xfnm[
  B.]}, \bibinfo{author}{Lundell\xfnm[ B.]}, \bibinfo{author}{O~Conchuir\xfnm[
  E.]}.
\newblock \bibinfo{title}{A framework for considering opportunities and threats
  in distributed software development}.
\newblock In: \bibinfo{booktitle}{Proceedings of the International Workshop on
  Distributed Software Development}. \bibinfo{year}{2005}. .
\bibitem[{Ahmed and Rafiq(2006)}]{Ahmed_2006}
\bibinfo{author}{Ahmed\xfnm[ P.K.]}, \bibinfo{author}{Rafiq\xfnm[ M.]}.
\newblock \bibinfo{title}{Internal Marketing: Tools and Concepts for
  Customer-Focused Management}.
\newblock \bibinfo{address}{Oxford}: \bibinfo{publisher}{Butterworth-Heineman},
  \bibinfo{year}{2006}.
\newblock \DOIprefix\doi{10.4324/9780080509037}.
\bibitem[{Al~Shraah et~al.(2022)Al~Shraah, Abu-Rumman, Al~Madi, Alhammad and
  AlJboor}]{Al_2022}
\bibinfo{author}{Al~Shraah\xfnm[ A.]}, \bibinfo{author}{Abu-Rumman\xfnm[ A.]},
  \bibinfo{author}{Al~Madi\xfnm[ F.]}, \bibinfo{author}{Alhammad\xfnm[
  F.A.F.]}, \bibinfo{author}{AlJboor\xfnm[ A.A.]}.
\newblock \bibinfo{title}{The impact of quality management practices on
  knowledge management processes: a study of a social security corporation in
  jordan}.
\newblock \bibinfo{journal}{The TQM Journal}
  \bibinfo{year}{2022};\bibinfo{volume}{34}(\bibinfo{number}{4}):\bibinfo{pages}{605--626}.
\bibitem[{Alavi and Leidner(1999)}]{Alavi_1999}
\bibinfo{author}{Alavi\xfnm[ M.]}, \bibinfo{author}{Leidner\xfnm[ D.]}.
\newblock \bibinfo{title}{Knowledge {Management} {Systems}: {Issues},
  {Challenges}, and {Benefits}}.
\newblock \bibinfo{journal}{Communications of the Association for Information
  Systems} \bibinfo{year}{1999};\bibinfo{volume}{1}.
\newblock \DOIprefix\doi{10.17705/1CAIS.00107}.
\bibitem[{Alavi and Leidner(2001)}]{Alavi_2001}
\bibinfo{author}{Alavi\xfnm[ M.]}, \bibinfo{author}{Leidner\xfnm[ D.E.]}.
\newblock \bibinfo{title}{Knowledge management and knowledge management
  systems: Conceptual foundations and research issues}.
\newblock \bibinfo{journal}{MIS quarterly}
  \bibinfo{year}{2001};:\bibinfo{pages}{107--136}\DOIprefix\doi{10.2307/3250961}.
\bibitem[{Alavi and Tiwana(2002)}]{Alavi_2002}
\bibinfo{author}{Alavi\xfnm[ M.]}, \bibinfo{author}{Tiwana\xfnm[ A.]}.
\newblock \bibinfo{title}{Knowledge integration in virtual teams: The potential
  role of {KMS}}.
\newblock \bibinfo{journal}{Journal of the American Society for Information
  Science and Technology}
  \bibinfo{year}{2002};\bibinfo{volume}{53}(\bibinfo{number}{12}):\bibinfo{pages}{1029--1037}.
\newblock \DOIprefix\doi{10.1002/asi.10107}.
\bibitem[{Almeida and Soares(2014)}]{Almeida_2014}
\bibinfo{author}{Almeida\xfnm[ M.V.]}, \bibinfo{author}{Soares\xfnm[ A.L.]}.
\newblock \bibinfo{title}{Knowledge sharing in project-based organizations:
  Overcoming the informational limbo}.
\newblock \bibinfo{journal}{International Journal of Information Management}
  \bibinfo{year}{2014};\bibinfo{volume}{34}(\bibinfo{number}{6}):\bibinfo{pages}{770--779}.
\newblock \DOIprefix\doi{10.1016/j.ijinfomgt.2014.07.003}.
\bibitem[{Andreeva and Kianto(2012)}]{Andreeva_2012}
\bibinfo{author}{Andreeva\xfnm[ T.]}, \bibinfo{author}{Kianto\xfnm[ A.]}.
\newblock \bibinfo{title}{Does knowledge management really matter? linking
  knowledge management practices, competitiveness and economic performance}.
\newblock \bibinfo{journal}{Journal of knowledge management}
  \bibinfo{year}{2012};\bibinfo{volume}{16}(\bibinfo{number}{4}):\bibinfo{pages}{617--636}.
\newblock \DOIprefix\doi{10.1108/13673271211246185}.
\bibitem[{Asp et~al.(2021)Asp, Taipalus and Seppänen}]{Asp_2021}
\bibinfo{author}{Asp\xfnm[ J.]}, \bibinfo{author}{Taipalus\xfnm[ T.]},
  \bibinfo{author}{Seppänen\xfnm[ V.]}.
\newblock \bibinfo{title}{Challenges in geographically distributed information
  system development: A case study}.
\newblock In: \bibinfo{booktitle}{2021 IEEE 45th Annual Computers, Software,
  and Applications Conference (COMPSAC)}. \bibinfo{year}{2021}. p.
  \bibinfo{pages}{452--458}.
\newblock \DOIprefix\doi{10.1109/COMPSAC51774.2021.00069}.
\bibitem[{Aurum et~al.(2008)Aurum, Daneshgar and Ward}]{Aurum_2008}
\bibinfo{author}{Aurum\xfnm[ A.]}, \bibinfo{author}{Daneshgar\xfnm[ F.]},
  \bibinfo{author}{Ward\xfnm[ J.]}.
\newblock \bibinfo{title}{Investigating knowledge management practices in
  software development organisations–an australian experience}.
\newblock \bibinfo{journal}{Information and Software Technology}
  \bibinfo{year}{2008};\bibinfo{volume}{50}(\bibinfo{number}{6}):\bibinfo{pages}{511--533}.
\newblock \DOIprefix\doi{10.1016/j.infsof.2007.05.005}.
\bibitem[{Bailey and Clarke(2001)}]{Bailey_2001}
\bibinfo{author}{Bailey\xfnm[ C.]}, \bibinfo{author}{Clarke\xfnm[ M.]}.
\newblock \bibinfo{title}{Managing knowledge for personal and organisational
  benefit}.
\newblock \bibinfo{journal}{Journal of Knowledge Management}
  \bibinfo{year}{2001};\DOIprefix\doi{10.1108/13673270110384400}.
\bibitem[{Bartol and Srivastava(2002)}]{Bartol_2002}
\bibinfo{author}{Bartol\xfnm[ K.M.]}, \bibinfo{author}{Srivastava\xfnm[ A.]}.
\newblock \bibinfo{title}{Encouraging knowledge sharing: The role of
  organizational reward systems}.
\newblock \bibinfo{journal}{Journal of Leadership \& Organizational Studies}
  \bibinfo{year}{2002};\bibinfo{volume}{9}(\bibinfo{number}{1}):\bibinfo{pages}{64--76}.
\newblock \DOIprefix\doi{10.1177/107179190200900105}.
\bibitem[{Battin et~al.(2001)Battin, Crocker, Kreidler and
  Subramanian}]{Battin_2001}
\bibinfo{author}{Battin\xfnm[ R.D.]}, \bibinfo{author}{Crocker\xfnm[ R.]},
  \bibinfo{author}{Kreidler\xfnm[ J.]}, \bibinfo{author}{Subramanian\xfnm[
  K.]}.
\newblock \bibinfo{title}{Leveraging resources in global software development}.
\newblock \bibinfo{journal}{IEEE Software}
  \bibinfo{year}{2001};\bibinfo{volume}{18}(\bibinfo{number}{2}):\bibinfo{pages}{70--77}.
\newblock \DOIprefix\doi{10.1109/52.914750}.
\bibitem[{Beck et~al.(2001)Beck, Beedle, van Bennekum, Cockburn, Cunningham,
  Fowler, Grenning, Highsmith, Hunt, Jeffries, Kern, Marick, Martin, Mellor,
  Schwaber, Sutherland and Thomas}]{Beck_2001}
\bibinfo{author}{Beck\xfnm[ K.]}, \bibinfo{author}{Beedle\xfnm[ M.]},
  \bibinfo{author}{van Bennekum\xfnm[ A.]}, \bibinfo{author}{Cockburn\xfnm[
  A.]}, \bibinfo{author}{Cunningham\xfnm[ W.]}, \bibinfo{author}{Fowler\xfnm[
  M.]}, \bibinfo{author}{Grenning\xfnm[ J.]}, \bibinfo{author}{Highsmith\xfnm[
  J.]}, \bibinfo{author}{Hunt\xfnm[ A.]}, \bibinfo{author}{Jeffries\xfnm[ R.]},
  \bibinfo{author}{Kern\xfnm[ J.]}, \bibinfo{author}{Marick\xfnm[ B.]},
  \bibinfo{author}{Martin\xfnm[ R.C.]}, \bibinfo{author}{Mellor\xfnm[ S.]},
  \bibinfo{author}{Schwaber\xfnm[ K.]}, \bibinfo{author}{Sutherland\xfnm[ J.]},
  \bibinfo{author}{Thomas\xfnm[ D.]}.
\newblock \bibinfo{title}{Manifesto for agile software development}.
\newblock \bibinfo{howpublished}{Retrieved from
  \url{https://www.agilemanifesto.org}}; \bibinfo{year}{2001}.
\bibitem[{Andreia~de Bem~Machado and Lanzalonga(2022)}]{Machado_2022}
\bibinfo{author}{Andreia~de Bem~Machado Silvana~Secinaro\xfnm[ D.C.]},
  \bibinfo{author}{Lanzalonga\xfnm[ F.]}.
\newblock \bibinfo{title}{Knowledge management and digital transformation for
  industry 4.0: a structured literature review}.
\newblock \bibinfo{journal}{Knowledge Management Research \& Practice}
  \bibinfo{year}{2022};\bibinfo{volume}{20}(\bibinfo{number}{2}):\bibinfo{pages}{320--338}.
\newblock \URLprefix \url{https://doi.org/10.1080/14778238.2021.2015261}.
  \DOIprefix\doi{10.1080/14778238.2021.2015261}.
  \href{http://arxiv.org/abs/https://doi.org/10.1080/14778238.2021.2015261}{\tt
  arXiv:https://doi.org/10.1080/14778238.2021.2015261}.
\bibitem[{Bock et~al.(2005)Bock, Zmud, Kim and Lee}]{Bock_2005}
\bibinfo{author}{Bock\xfnm[ G.]}, \bibinfo{author}{Zmud\xfnm[ R.W.]},
  \bibinfo{author}{Kim\xfnm[ Y.]}, \bibinfo{author}{Lee\xfnm[ J.N.]}.
\newblock \bibinfo{title}{Behavioral intention formation in knowledge sharing:
  Examining the roles of extrinsic motivators, social-psychological forces, and
  organizational climate}.
\newblock \bibinfo{journal}{MIS quarterly}
  \bibinfo{year}{2005};\bibinfo{volume}{29}(\bibinfo{number}{1}):\bibinfo{pages}{87--111}.
\newblock \DOIprefix\doi{10.2307/25148669}.
\bibitem[{Bock and Kim(2002)}]{Bock_2002}
\bibinfo{author}{Bock\xfnm[ G.W.]}, \bibinfo{author}{Kim\xfnm[ Y.G.]}.
\newblock \bibinfo{title}{Breaking the myths of rewards: An exploratory study
  of attitudes about knowledge sharing}.
\newblock \bibinfo{journal}{Information Resources Management Journal (IRMJ)}
  \bibinfo{year}{2002};\bibinfo{volume}{15}(\bibinfo{number}{2}):\bibinfo{pages}{14--21}.
\newblock \DOIprefix\doi{10.4018/irmj.2002040102}.
\bibitem[{Bolisani et~al.(2023)Bolisani, Scarso and Kassaneh}]{Bolisani_2023}
\bibinfo{author}{Bolisani\xfnm[ E.]}, \bibinfo{author}{Scarso\xfnm[ E.]},
  \bibinfo{author}{Kassaneh\xfnm[ T.C.]}.
\newblock \bibinfo{title}{The pervasive identity of knowledge management:
  Consolidation or dilution?}
\newblock In: \bibinfo{booktitle}{The Future of Knowledge Management:
  Reflections from the 10th Anniversary of the International Association of
  Knowledge Management (IAKM)}. \bibinfo{publisher}{Springer};
  \bibinfo{year}{2023}. p. \bibinfo{pages}{23--45}.
\bibitem[{Braganza(2004)}]{Braganza_2004}
\bibinfo{author}{Braganza\xfnm[ A.]}.
\newblock \bibinfo{title}{Rethinking the data--information--knowledge
  hierarchy: Towards a case-based model}.
\newblock \bibinfo{journal}{International Journal of Information Management}
  \bibinfo{year}{2004};\bibinfo{volume}{24}(\bibinfo{number}{4}):\bibinfo{pages}{347--356}.
\newblock \DOIprefix\doi{10.1016/j.ijinfomgt.2004.04.007}.
\bibitem[{Butt et~al.(2019)Butt, Nawaz, Hussain, Sousa, Wang, Sumbal and
  Shujahat}]{Butt_2019}
\bibinfo{author}{Butt\xfnm[ M.A.]}, \bibinfo{author}{Nawaz\xfnm[ F.]},
  \bibinfo{author}{Hussain\xfnm[ S.]}, \bibinfo{author}{Sousa\xfnm[ M.J.]},
  \bibinfo{author}{Wang\xfnm[ M.]}, \bibinfo{author}{Sumbal\xfnm[ M.S.]},
  \bibinfo{author}{Shujahat\xfnm[ M.]}.
\newblock \bibinfo{title}{Individual knowledge management engagement,
  knowledge-worker productivity, and innovation performance in knowledge-based
  organizations: the implications for knowledge processes and knowledge-based
  systems}.
\newblock \bibinfo{journal}{Computational and Mathematical Organization Theory}
  \bibinfo{year}{2019};\bibinfo{volume}{25}:\bibinfo{pages}{336--356}.
\bibitem[{Choi et~al.(2010)Choi, Lee and Yoo}]{Choi_2010}
\bibinfo{author}{Choi\xfnm[ S.Y.]}, \bibinfo{author}{Lee\xfnm[ H.]},
  \bibinfo{author}{Yoo\xfnm[ Y.]}.
\newblock \bibinfo{title}{The impact of information technology and transactive
  memory systems on knowledge sharing, application, and team performance: A
  field study}.
\newblock \bibinfo{journal}{MIS Quarterly}
  \bibinfo{year}{2010};\bibinfo{volume}{34}(\bibinfo{number}{4}):\bibinfo{pages}{855--870}.
\newblock \DOIprefix\doi{10.2307/25750708}.
\bibitem[{Connelly and Kelloway(2003)}]{Connelly_2003}
\bibinfo{author}{Connelly\xfnm[ C.E.]}, \bibinfo{author}{Kelloway\xfnm[ E.K.]}.
\newblock \bibinfo{title}{Predictors of employees’ perceptions of knowledge
  sharing cultures}.
\newblock \bibinfo{journal}{Leadership \& Organization Development Journal}
  \bibinfo{year}{2003};\DOIprefix\doi{10.1108/01437730310485815}.
\bibitem[{De~Clercq et~al.(2013)De~Clercq, Thongpapanl and
  Dimov}]{De_Clercq_2013}
\bibinfo{author}{De~Clercq\xfnm[ D.]}, \bibinfo{author}{Thongpapanl\xfnm[ N.]},
  \bibinfo{author}{Dimov\xfnm[ D.]}.
\newblock \bibinfo{title}{Organizational social capital, formalization, and
  internal knowledge sharing in entrepreneurial orientation formation}.
\newblock \bibinfo{journal}{Entrepreneurship Theory and Practice}
  \bibinfo{year}{2013};\bibinfo{volume}{37}(\bibinfo{number}{3}):\bibinfo{pages}{505--537}.
\newblock \DOIprefix\doi{10.1111/etap.12021}.
\bibitem[{De~Gooijer(2000)}]{De_Gooijer_2000}
\bibinfo{author}{De~Gooijer\xfnm[ J.]}.
\newblock \bibinfo{title}{Designing a knowledge management performance
  framework}.
\newblock \bibinfo{journal}{Journal of knowledge management}
  \bibinfo{year}{2000};\DOIprefix\doi{10.1108/13673270010379858}.
\bibitem[{Desouza et~al.(2006)Desouza, Awazu and Baloh}]{Desouza_2006}
\bibinfo{author}{Desouza\xfnm[ K.C.]}, \bibinfo{author}{Awazu\xfnm[ Y.]},
  \bibinfo{author}{Baloh\xfnm[ P.]}.
\newblock \bibinfo{title}{Managing knowledge in global software development
  efforts: Issues and practices}.
\newblock \bibinfo{journal}{IEEE Software}
  \bibinfo{year}{2006};\bibinfo{volume}{23}(\bibinfo{number}{5}):\bibinfo{pages}{30--37}.
\newblock \DOIprefix\doi{10.1109/ms.2006.135}.
\bibitem[{Dingsøyr et~al.(2009)Dingsøyr, Bjørnson and Shull}]{Dingsoyr_2009}
\bibinfo{author}{Dingsøyr\xfnm[ T.]}, \bibinfo{author}{Bjørnson\xfnm[ F.O.]},
  \bibinfo{author}{Shull\xfnm[ F.]}.
\newblock \bibinfo{title}{What do we know about knowledge management? practical
  implications for software engineering}.
\newblock \bibinfo{journal}{IEEE software}
  \bibinfo{year}{2009};\bibinfo{volume}{26}(\bibinfo{number}{3}):\bibinfo{pages}{100--103}.
\newblock \DOIprefix\doi{10.1109/ms.2009.82}.
\bibitem[{Dingsøyr and Smite(2013)}]{Dingsoyr_2013}
\bibinfo{author}{Dingsøyr\xfnm[ T.]}, \bibinfo{author}{Smite\xfnm[ D.]}.
\newblock \bibinfo{title}{Managing knowledge in global software development
  projects}.
\newblock \bibinfo{journal}{IT Professional}
  \bibinfo{year}{2013};\bibinfo{volume}{16}(\bibinfo{number}{1}):\bibinfo{pages}{22--29}.
\bibitem[{Dorairaj et~al.(2012)Dorairaj, Noble and Malik}]{Dorairaj_2012}
\bibinfo{author}{Dorairaj\xfnm[ S.]}, \bibinfo{author}{Noble\xfnm[ J.]},
  \bibinfo{author}{Malik\xfnm[ P.]}.
\newblock \bibinfo{title}{Knowledge management in distributed agile software
  development}.
\newblock In: \bibinfo{booktitle}{2012 Agile Conference}.
  \bibinfo{organization}{IEEE}; \bibinfo{year}{2012}. p.
  \bibinfo{pages}{64--73}.
\newblock \DOIprefix\doi{10.1109/agile.2012.17}.
\bibitem[{Ebert and De~Neve(2001)}]{Ebert_2001}
\bibinfo{author}{Ebert\xfnm[ C.]}, \bibinfo{author}{De~Neve\xfnm[ P.]}.
\newblock \bibinfo{title}{Surviving global software development}.
\newblock \bibinfo{journal}{IEEE software}
  \bibinfo{year}{2001};\bibinfo{volume}{18}(\bibinfo{number}{2}):\bibinfo{pages}{62--69}.
\bibitem[{Faraj and Sproull(2000)}]{Faraj_2000}
\bibinfo{author}{Faraj\xfnm[ S.]}, \bibinfo{author}{Sproull\xfnm[ L.]}.
\newblock \bibinfo{title}{Coordinating expertise in software development
  teams}.
\newblock \bibinfo{journal}{Management science}
  \bibinfo{year}{2000};\bibinfo{volume}{46}(\bibinfo{number}{12}):\bibinfo{pages}{1554--1568}.
\newblock \DOIprefix\doi{10.1287/mnsc.46.12.1554.12072}.
\bibitem[{Fischer and Otswald(2001)}]{Fischer_2001}
\bibinfo{author}{Fischer\xfnm[ G.]}, \bibinfo{author}{Otswald\xfnm[ J.]}.
\newblock \bibinfo{title}{Knowledge management: problems, promises, realities,
  and challenges}.
\newblock \bibinfo{journal}{IEEE Intelligent Systems}
  \bibinfo{year}{2001};\bibinfo{volume}{16}(\bibinfo{number}{1}):\bibinfo{pages}{60--72}.
\newblock \DOIprefix\doi{10.1109/5254.912386}.
\bibitem[{Foress and Camison(2016)}]{Fores_2016}
\bibinfo{author}{Foress\xfnm[ B.]}, \bibinfo{author}{Camison\xfnm[ C.]}.
\newblock \bibinfo{title}{Does incremental and radical innovation performance
  depend on different types of knowledge accumulation capabilities and
  organizational size?}
\newblock \bibinfo{journal}{Journal of business research}
  \bibinfo{year}{2016};\bibinfo{volume}{69}(\bibinfo{number}{2}):\bibinfo{pages}{831--848}.
\newblock \DOIprefix\doi{10.1016/j.jbusres.2015.07.006}.
\bibitem[{Gold et~al.(2001)Gold, Malhotra and Segard}]{Gold_2001}
\bibinfo{author}{Gold\xfnm[ A.]}, \bibinfo{author}{Malhotra\xfnm[ A.]},
  \bibinfo{author}{Segard\xfnm[ A.]}.
\newblock \bibinfo{title}{Knowledge management: an organizational capabilities
  perspective}.
\newblock \bibinfo{journal}{Journal of Management Information Systems}
  \bibinfo{year}{2001};\bibinfo{volume}{18}(\bibinfo{number}{1}):\bibinfo{pages}{185--214}.
\newblock \DOIprefix\doi{10.1080/07421222.2001.11045669}.
\bibitem[{Gressgard(2015)}]{Gressgard_2015}
\bibinfo{author}{Gressgard\xfnm[ L.J.]}.
\newblock \bibinfo{title}{Antecedents of knowledge exchange systems usage:
  Motivational and work environment factors}.
\newblock \bibinfo{journal}{Knowledge and Process Management}
  \bibinfo{year}{2015};\bibinfo{volume}{22}(\bibinfo{number}{2}):\bibinfo{pages}{112--125}.
\newblock \DOIprefix\doi{10.1002/kpm.1471}.
\bibitem[{Griffith et~al.(2003)Griffith, Sawyer and Neale}]{Griffith_2003}
\bibinfo{author}{Griffith\xfnm[ T.L.]}, \bibinfo{author}{Sawyer\xfnm[ J.E.]},
  \bibinfo{author}{Neale\xfnm[ M.A.]}.
\newblock \bibinfo{title}{Virtualness and knowledge in teams: managing the love
  triangle of organizations, individuals, and information technology}.
\newblock \bibinfo{journal}{MIS Quarterly}
  \bibinfo{year}{2003};\bibinfo{volume}{27}(\bibinfo{number}{2}):\bibinfo{pages}{265--287}.
\newblock \DOIprefix\doi{10.2307/30036531}.
\bibitem[{Gunnlaugsdottir(2003)}]{Gunnlaugsdottir_2003}
\bibinfo{author}{Gunnlaugsdottir\xfnm[ J.]}.
\newblock \bibinfo{title}{Seek and you will find, share and you will benefit:
  organising knowledge using groupware systems}.
\newblock \bibinfo{journal}{International Journal of Information Management}
  \bibinfo{year}{2003};\bibinfo{volume}{23}(\bibinfo{number}{5}):\bibinfo{pages}{363--380}.
\newblock \DOIprefix\doi{10.1016/s0268-4012(03)00064-1}.
\bibitem[{Halil~Zaim and Tarim(2019)}]{Zaim_2019}
\bibinfo{author}{Halil~Zaim\xfnm[ S.M.]}, \bibinfo{author}{Tarim\xfnm[ M.]}.
\newblock \bibinfo{title}{Relationship between knowledge management processes
  and performance: critical role of knowledge utilization in organizations}.
\newblock \bibinfo{journal}{Knowledge Management Research \& Practice}
  \bibinfo{year}{2019};\bibinfo{volume}{17}(\bibinfo{number}{1}):\bibinfo{pages}{24--38}.
\newblock \URLprefix \url{https://doi.org/10.1080/14778238.2018.1538669}.
  \DOIprefix\doi{10.1080/14778238.2018.1538669}.
  \href{http://arxiv.org/abs/https://doi.org/10.1080/14778238.2018.1538669}{\tt
  arXiv:https://doi.org/10.1080/14778238.2018.1538669}.
\bibitem[{Hansen et~al.(1999)Hansen, Nohria and Tierney}]{Hansen_1999}
\bibinfo{author}{Hansen\xfnm[ M.T.]}, \bibinfo{author}{Nohria\xfnm[ N.]},
  \bibinfo{author}{Tierney\xfnm[ T.]}.
\newblock \bibinfo{title}{What's your strategy for managing knowledge?}
\newblock \bibinfo{journal}{Harvard Business Review}
  \bibinfo{year}{1999};\bibinfo{volume}{77}(\bibinfo{number}{2}):\bibinfo{pages}{106--116}.
\newblock \DOIprefix\doi{10.4324/9780080941042-9}.
\bibitem[{Asrar-ul Haq and Anwar(2016)}]{Asrar_2016}
\bibinfo{author}{Asrar-ul Haq\xfnm[ M.]}, \bibinfo{author}{Anwar\xfnm[ S.]}.
\newblock \bibinfo{title}{A systematic review of knowledge management and
  knowledge sharing: {Trends}, issues, and challenges}.
\newblock \bibinfo{journal}{Cogent Business \& Management}
  \bibinfo{year}{2016};\bibinfo{volume}{3}(\bibinfo{number}{1}):\bibinfo{pages}{1127744}.
\newblock \DOIprefix\doi{10.1080/23311975.2015.1127744}.
\bibitem[{Hau et~al.(2013)Hau, Kim, Lee and Kim}]{Hau_2013}
\bibinfo{author}{Hau\xfnm[ Y.S.]}, \bibinfo{author}{Kim\xfnm[ B.]},
  \bibinfo{author}{Lee\xfnm[ H.]}, \bibinfo{author}{Kim\xfnm[ Y.]}.
\newblock \bibinfo{title}{The effects of individual motivations and social
  capital on employees’ tacit and explicit knowledge sharing intentions}.
\newblock \bibinfo{journal}{International Journal of Information Management}
  \bibinfo{year}{2013};\bibinfo{volume}{33}(\bibinfo{number}{2}):\bibinfo{pages}{356--366}.
\newblock \DOIprefix\doi{10.1016/j.ijinfomgt.2012.10.009}.
\bibitem[{Herbsleb et~al.(2001)Herbsleb, Mockus, Finholt and
  Grinter}]{Herbsleb_2001}
\bibinfo{author}{Herbsleb\xfnm[ J.D.]}, \bibinfo{author}{Mockus\xfnm[ A.]},
  \bibinfo{author}{Finholt\xfnm[ T.A.]}, \bibinfo{author}{Grinter\xfnm[ R.E.]}.
\newblock \bibinfo{title}{An empirical study of global software development:
  distance and speed}.
\newblock In: \bibinfo{booktitle}{Proceedings of the 23rd international
  conference on software engineering}. \bibinfo{organization}{IEEE Computer
  Society}; \bibinfo{year}{2001}. p. \bibinfo{pages}{81--90}.
\newblock \DOIprefix\doi{10.1109/icse.2001.919083}.
\bibitem[{Herbsleb and Moitra(2001)}]{Herbsleb_2001b}
\bibinfo{author}{Herbsleb\xfnm[ J.D.]}, \bibinfo{author}{Moitra\xfnm[ D.]}.
\newblock \bibinfo{title}{Global software development}.
\newblock \bibinfo{journal}{IEEE software}
  \bibinfo{year}{2001};\bibinfo{volume}{18}(\bibinfo{number}{2}):\bibinfo{pages}{16--20}.
\bibitem[{Hsieh and Shannon(2005)}]{Hsieh_2005}
\bibinfo{author}{Hsieh\xfnm[ H.F.]}, \bibinfo{author}{Shannon\xfnm[ S.E.]}.
\newblock \bibinfo{title}{Three approaches to qualitative content analysis}.
\newblock \bibinfo{journal}{Qualitative health research}
  \bibinfo{year}{2005};\bibinfo{volume}{15}(\bibinfo{number}{9}):\bibinfo{pages}{1277--1288}.
\newblock \DOIprefix\doi{10.1177/1049732305276687}.
\bibitem[{Hume and Hume(2015)}]{Hume_2015}
\bibinfo{author}{Hume\xfnm[ C.]}, \bibinfo{author}{Hume\xfnm[ M.]}.
\newblock \bibinfo{title}{The critical role of internal marketing in knowledge
  management in not-for-profit organizations}.
\newblock \bibinfo{journal}{Journal of Nonprofit \& Public Sector Marketing}
  \bibinfo{year}{2015};\bibinfo{volume}{27}:\bibinfo{pages}{23--47}.
\newblock \DOIprefix\doi{10.1080/10495142.2014.934567}.
\bibitem[{{International Organization for Standardization}(2018)}]{iso30401}
\bibinfo{author}{{International Organization for Standardization}\xfnm[]}.
\newblock \bibinfo{title}{{ISO 30401:2018} knowledge management systems --
  requirements}.
\newblock \bibinfo{year}{2018}.
\bibitem[{Jackson et~al.(2006)Jackson, Chuang, Harden, Jiang and
  Joseph}]{Jackson_2006}
\bibinfo{author}{Jackson\xfnm[ S.E.]}, \bibinfo{author}{Chuang\xfnm[ C.H.]},
  \bibinfo{author}{Harden\xfnm[ E.E.]}, \bibinfo{author}{Jiang\xfnm[ Y.]},
  \bibinfo{author}{Joseph\xfnm[ J.M.]}.
\newblock \bibinfo{title}{Toward developing human resource management systems
  for knowledge-intensive teamwork}.
\newblock \bibinfo{journal}{Research in Personnel and Human Resources
  Management}
  \bibinfo{year}{2006};\bibinfo{volume}{25}(\bibinfo{number}{6}):\bibinfo{pages}{27--70}.
\newblock \DOIprefix\doi{10.1016/s0742-7301(06)25002-3}.
\bibitem[{Kalkan(2008)}]{Kalkan_2008}
\bibinfo{author}{Kalkan\xfnm[ V.D.]}.
\newblock \bibinfo{title}{An overall view of knowledge management challenges
  for global business}.
\newblock \bibinfo{journal}{Business Process Management Journal}
  \bibinfo{year}{2008};\bibinfo{volume}{14}(\bibinfo{number}{3}):\bibinfo{pages}{390--400}.
\bibitem[{Kim and Lee(2006)}]{Kim_2006}
\bibinfo{author}{Kim\xfnm[ S.]}, \bibinfo{author}{Lee\xfnm[ H.]}.
\newblock \bibinfo{title}{The impact of organizational context and information
  technology on employee knowledge-sharing capabilities}.
\newblock \bibinfo{journal}{Public Administration Review}
  \bibinfo{year}{2006};\bibinfo{volume}{66}(\bibinfo{number}{3}):\bibinfo{pages}{370--385}.
\newblock \DOIprefix\doi{10.1111/j.1540-6210.2006.00595.x}.
\bibitem[{Klitmøller and Lauring(2013)}]{Klitmoller_2013}
\bibinfo{author}{Klitmøller\xfnm[ A.]}, \bibinfo{author}{Lauring\xfnm[ J.]}.
\newblock \bibinfo{title}{When global virtual teams share knowledge: Media
  richness, cultural difference and language commonality}.
\newblock \bibinfo{journal}{Journal of World Business: JWB}
  \bibinfo{year}{2013};\bibinfo{volume}{48}(\bibinfo{number}{3}):\bibinfo{pages}{398--406}.
\newblock \DOIprefix\doi{10.1016/j.jwb.2012.07.023}.
\bibitem[{Kotlarsky and Oshri(2005)}]{Kotlarsky_2005}
\bibinfo{author}{Kotlarsky\xfnm[ J.]}, \bibinfo{author}{Oshri\xfnm[ I.]}.
\newblock \bibinfo{title}{Social ties, knowledge sharing and successful
  collaboration in globally distributed system development projects}.
\newblock \bibinfo{journal}{European Journal of Information Systems}
  \bibinfo{year}{2005};\bibinfo{volume}{14}(\bibinfo{number}{1}):\bibinfo{pages}{37–48}.
\newblock \DOIprefix\doi{10.1057/palgrave.ejis.3000520}.
\bibitem[{Lee and Choi(2003)}]{Lee_2003}
\bibinfo{author}{Lee\xfnm[ H.]}, \bibinfo{author}{Choi\xfnm[ B.]}.
\newblock \bibinfo{title}{Knowledge management enablers, processes, and
  organizational performance: An integrative view and empirical examination}.
\newblock \bibinfo{journal}{Journal of Management Information Systems}
  \bibinfo{year}{2003};\bibinfo{volume}{20}(\bibinfo{number}{1}):\bibinfo{pages}{179--228}.
\newblock \DOIprefix\doi{10.1080/07421222.2003.11045756}.
\bibitem[{Leiblein and Madsen(2009)}]{Leiblein_2009}
\bibinfo{author}{Leiblein\xfnm[ M.J.]}, \bibinfo{author}{Madsen\xfnm[ T.L.]}.
\newblock \bibinfo{title}{Unbundling competitive heterogeneity: Incentive
  structures and capability influences on technological innovation}.
\newblock \bibinfo{journal}{Strategic Management Journal}
  \bibinfo{year}{2009};\bibinfo{volume}{30}(\bibinfo{number}{7}):\bibinfo{pages}{711--735}.
\newblock \DOIprefix\doi{10.2139/ssrn.1000078}.
\bibitem[{Lin(2008)}]{Lin_2008}
\bibinfo{author}{Lin\xfnm[ W.B.]}.
\newblock \bibinfo{title}{The effect of knowledge sharing model}.
\newblock \bibinfo{journal}{Expert Systems with Applications}
  \bibinfo{year}{2008};\bibinfo{volume}{34}(\bibinfo{number}{2}):\bibinfo{pages}{1508--1521}.
\newblock \DOIprefix\doi{10.1016/j.eswa.2007.01.015}.
\bibitem[{Manteli et~al.(2011)Manteli, Van Den~Hooff, Tang and
  Van~Vliet}]{Manteli_2011}
\bibinfo{author}{Manteli\xfnm[ C.]}, \bibinfo{author}{Van Den~Hooff\xfnm[ B.]},
  \bibinfo{author}{Tang\xfnm[ A.]}, \bibinfo{author}{Van~Vliet\xfnm[ H.]}.
\newblock \bibinfo{title}{The impact of multi-site software governance on
  knowledge management}.
\newblock In: \bibinfo{booktitle}{2011 IEEE Sixth International Conference on
  Global Software Engineering}. \bibinfo{organization}{IEEE};
  \bibinfo{year}{2011}. p. \bibinfo{pages}{40--49}.
\newblock \DOIprefix\doi{10.1109/icgse.2011.16}.
\bibitem[{Markus(2001)}]{Markus_2001}
\bibinfo{author}{Markus\xfnm[ M.L.]}.
\newblock \bibinfo{title}{Toward a theory of knowledge reuse: Types of
  knowledge reuse situations and factors in reuse success}.
\newblock \bibinfo{journal}{Journal of Management Information Systems}
  \bibinfo{year}{2001};\bibinfo{volume}{18}(\bibinfo{number}{1}):\bibinfo{pages}{57--93}.
\newblock \DOIprefix\doi{10.1080/07421222.2001.11045671}.
\bibitem[{Mazzei(2010)}]{Mazzei_2010}
\bibinfo{author}{Mazzei\xfnm[ A.]}.
\newblock \bibinfo{title}{Promoting active communication behaviours through
  internal communication}.
\newblock \bibinfo{journal}{Corporate Communications}
  \bibinfo{year}{2010};\bibinfo{volume}{15}(\bibinfo{number}{3}):\bibinfo{pages}{221--234}.
\newblock \DOIprefix\doi{10.1108/13563281011068096}.
\bibitem[{Nahapiet and Ghoshal(1998)}]{Nahapiet_1998}
\bibinfo{author}{Nahapiet\xfnm[ J.]}, \bibinfo{author}{Ghoshal\xfnm[ S.]}.
\newblock \bibinfo{title}{Social capital, intellectual capital, and the
  organizational advantage}.
\newblock \bibinfo{journal}{Academy of management review}
  \bibinfo{year}{1998};\bibinfo{volume}{23}(\bibinfo{number}{2}):\bibinfo{pages}{242--266}.
\newblock \DOIprefix\doi{10.1016/b978-0-7506-7222-1.50009-x}.
\bibitem[{Nakash and Bouhnik(2022)}]{Nakash_2022}
\bibinfo{author}{Nakash\xfnm[ M.]}, \bibinfo{author}{Bouhnik\xfnm[ D.]}.
\newblock \bibinfo{title}{Risks in the absence of optimal knowledge management
  in knowledge-intensive organizations}.
\newblock \bibinfo{journal}{VINE Journal of Information and Knowledge
  Management Systems}
  \bibinfo{year}{2022};\bibinfo{volume}{52}(\bibinfo{number}{1}):\bibinfo{pages}{87--101}.
\bibitem[{Nakash and Bouhnik(2023)}]{Nakash_2023}
\bibinfo{author}{Nakash\xfnm[ M.]}, \bibinfo{author}{Bouhnik\xfnm[ D.]}.
\newblock \bibinfo{title}{Challenges of justification of investment in
  organizational knowledge management}.
\newblock \bibinfo{journal}{Knowledge Management Research \& Practice}
  \bibinfo{year}{2023};\bibinfo{volume}{21}(\bibinfo{number}{4}):\bibinfo{pages}{703--713}.
\bibitem[{Niinimaki et~al.(2010)Niinimaki, Piri, Lassenius and
  Paasivaara}]{Niinimaki_2010}
\bibinfo{author}{Niinimaki\xfnm[ T.]}, \bibinfo{author}{Piri\xfnm[ A.]},
  \bibinfo{author}{Lassenius\xfnm[ C.]}, \bibinfo{author}{Paasivaara\xfnm[
  M.]}.
\newblock \bibinfo{title}{Reflecting the choice and usage of communication
  tools in gsd projects with media synchronicity theory}.
\newblock In: \bibinfo{booktitle}{2010 5th IEEE International Conference on
  Global Software Engineering}. \bibinfo{organization}{IEEE};
  \bibinfo{year}{2010}. p. \bibinfo{pages}{3--12}.
\newblock \DOIprefix\doi{10.1109/icgse.2010.11}.
\bibitem[{Noe et~al.(2003)Noe, Colquitt, Simmering and Alvarez}]{Noe_2003}
\bibinfo{author}{Noe\xfnm[ R.]}, \bibinfo{author}{Colquitt\xfnm[ J.]},
  \bibinfo{author}{Simmering\xfnm[ M.]}, \bibinfo{author}{Alvarez\xfnm[ S.]}.
\newblock \bibinfo{title}{Knowledge management: Developing intellectual and
  social capital}.
\newblock In: \bibinfo{editor}{Jackson\xfnm[ S.]}, \bibinfo{editor}{Hitt\xfnm[
  M.]}, \bibinfo{editor}{DeNisi\xfnm[ A.]}, editors.
  \bibinfo{booktitle}{Managing knowledge for sustained competitive advantage:
  Designing strategies for effective human resource management}.
  \bibinfo{address}{San Francisco}: \bibinfo{publisher}{Jossey-Bass};
  \bibinfo{year}{2003}. p. \bibinfo{pages}{209--242}.
\bibitem[{Nonaka and Takeuchi(1995)}]{Nonaka_1995}
\bibinfo{author}{Nonaka\xfnm[ I.]}, \bibinfo{author}{Takeuchi\xfnm[ H.]}.
\newblock \bibinfo{title}{The knowledge-creating company: How Japanese
  companies create the dynamics of innovation}.
\newblock \bibinfo{publisher}{Oxford university press}, \bibinfo{year}{1995}.
\newblock \DOIprefix\doi{10.1016/s0048-7333(97)80234-x}.
\bibitem[{Ode and Ayavoo(2020)}]{Ode_2020}
\bibinfo{author}{Ode\xfnm[ E.]}, \bibinfo{author}{Ayavoo\xfnm[ R.]}.
\newblock \bibinfo{title}{The mediating role of knowledge application in the
  relationship between knowledge management practices and firm innovation}.
\newblock \bibinfo{journal}{Journal of Innovation \& Knowledge}
  \bibinfo{year}{2020};\bibinfo{volume}{5}(\bibinfo{number}{3}):\bibinfo{pages}{210--218}.
\newblock \URLprefix
  \url{https://www.sciencedirect.com/science/article/pii/S2444569X19300423}.
  \DOIprefix\doi{https://doi.org/10.1016/j.jik.2019.08.002}.
\bibitem[{Oliva and Kotabe(2019)}]{Oliva_2019}
\bibinfo{author}{Oliva\xfnm[ F.L.]}, \bibinfo{author}{Kotabe\xfnm[ M.]}.
\newblock \bibinfo{title}{Barriers, practices, methods and knowledge management
  tools in startups}.
\newblock \bibinfo{journal}{Journal of knowledge management}
  \bibinfo{year}{2019};\bibinfo{volume}{23}(\bibinfo{number}{9}):\bibinfo{pages}{1838--1856}.
\bibitem[{Polanyi(1966)}]{Polanyi_1966}
\bibinfo{author}{Polanyi\xfnm[ M.]}.
\newblock \bibinfo{title}{The Tacit Dimension}.
\newblock \bibinfo{publisher}{Routledge}, \bibinfo{year}{1966}.
\newblock \DOIprefix\doi{10.1016/b978-0-7506-9718-7.50010-x}.
\bibitem[{Prikladnicki et~al.(2003)Prikladnicki, Nicolas~Audy and
  Evaristo}]{Prikladnicki_2003}
\bibinfo{author}{Prikladnicki\xfnm[ R.]}, \bibinfo{author}{Nicolas~Audy\xfnm[
  J.L.]}, \bibinfo{author}{Evaristo\xfnm[ R.]}.
\newblock \bibinfo{title}{Global software development in practice lessons
  learned}.
\newblock \bibinfo{journal}{Software Process: Improvement and Practice}
  \bibinfo{year}{2003};\bibinfo{volume}{8}(\bibinfo{number}{4}):\bibinfo{pages}{267--281}.
\newblock \DOIprefix\doi{10.1002/spip.188}.
\bibitem[{Razzaq et~al.(2019)Razzaq, Shujahat, Hussain, Nawaz, Wang, Ali and
  Tehseen}]{Razzaq_2019}
\bibinfo{author}{Razzaq\xfnm[ S.]}, \bibinfo{author}{Shujahat\xfnm[ M.]},
  \bibinfo{author}{Hussain\xfnm[ S.]}, \bibinfo{author}{Nawaz\xfnm[ F.]},
  \bibinfo{author}{Wang\xfnm[ M.]}, \bibinfo{author}{Ali\xfnm[ M.]},
  \bibinfo{author}{Tehseen\xfnm[ S.]}.
\newblock \bibinfo{title}{Knowledge management, organizational commitment and
  knowledge-worker performance: The neglected role of knowledge management in
  the public sector}.
\newblock \bibinfo{journal}{Business process management journal}
  \bibinfo{year}{2019};\bibinfo{volume}{25}(\bibinfo{number}{5}):\bibinfo{pages}{923--947}.
\bibitem[{Rus and Lindvall(2002)}]{Rus_2002}
\bibinfo{author}{Rus\xfnm[ I.]}, \bibinfo{author}{Lindvall\xfnm[ M.]}.
\newblock \bibinfo{title}{Knowledge management in software engineering}.
\newblock \bibinfo{journal}{IEEE Software}
  \bibinfo{year}{2002};\bibinfo{volume}{19}(\bibinfo{number}{3}):\bibinfo{pages}{26}.
\bibitem[{Ružić and Benazić(2021)}]{Ruzic_2021}
\bibinfo{author}{Ružić\xfnm[ E.]}, \bibinfo{author}{Benazić\xfnm[ D.]}.
\newblock \bibinfo{title}{The impact of internal knowledge sharing on sales
  department’s innovativeness and new product commercialization}.
\newblock \bibinfo{journal}{Organizacija}
  \bibinfo{year}{2021};\bibinfo{volume}{54}(\bibinfo{number}{2}):\bibinfo{pages}{147--160}.
\newblock \DOIprefix\doi{10.2478/orga-2021-0010}.
\bibitem[{Ryan and O'Connor(2009)}]{Ryan_2009}
\bibinfo{author}{Ryan\xfnm[ S.]}, \bibinfo{author}{O'Connor\xfnm[ R.V.]}.
\newblock \bibinfo{title}{Development of a team measure for tacit knowledge in
  software development teams}.
\newblock \bibinfo{journal}{Journal of Systems and Software}
  \bibinfo{year}{2009};\bibinfo{volume}{82}(\bibinfo{number}{2}):\bibinfo{pages}{229--240}.
\newblock \DOIprefix\doi{10.1016/j.jss.2008.05.037}.
\bibitem[{Ryan and O'Connor(2013)}]{Ryan_2013}
\bibinfo{author}{Ryan\xfnm[ S.]}, \bibinfo{author}{O'Connor\xfnm[ R.V.]}.
\newblock \bibinfo{title}{Acquiring and sharing tacit knowledge in software
  development teams: An empirical study}.
\newblock \bibinfo{journal}{Information and Software Technology}
  \bibinfo{year}{2013};\bibinfo{volume}{55}(\bibinfo{number}{9}):\bibinfo{pages}{1614--1624}.
\newblock \DOIprefix\doi{10.1016/j.infsof.2013.02.013}.
\bibitem[{Seba et~al.(2012)Seba, Rowley and Lambert}]{Seba_2012}
\bibinfo{author}{Seba\xfnm[ I.]}, \bibinfo{author}{Rowley\xfnm[ J.]},
  \bibinfo{author}{Lambert\xfnm[ S.]}.
\newblock \bibinfo{title}{Factors affecting attitudes and intentions towards
  knowledge sharing in the dubai police force}.
\newblock \bibinfo{journal}{International Journal of Information Management}
  \bibinfo{year}{2012};\bibinfo{volume}{32}(\bibinfo{number}{4}):\bibinfo{pages}{372--380}.
\newblock \DOIprefix\doi{10.1016/j.ijinfomgt.2011.12.003}.
\bibitem[{Serenko et~al.(2007)Serenko, Bontis and Hardie}]{Serenko_2007}
\bibinfo{author}{Serenko\xfnm[ A.]}, \bibinfo{author}{Bontis\xfnm[ N.]},
  \bibinfo{author}{Hardie\xfnm[ T.]}.
\newblock \bibinfo{title}{Organizational size and knowledge flow: a proposed
  theoretical link}.
\newblock \bibinfo{journal}{Journal of Intellectual Capital}
  \bibinfo{year}{2007};\bibinfo{volume}{8}(\bibinfo{number}{4}):\bibinfo{pages}{610--627}.
\newblock \DOIprefix\doi{10.1108/14691930710830783}.
\bibitem[{Sharp(2003)}]{Sharp_2003}
\bibinfo{author}{Sharp\xfnm[ D.]}.
\newblock \bibinfo{title}{Knowledge {Management} {Today}: {Challenges} and
  {Opportunities}}.
\newblock \bibinfo{journal}{Information Systems Management}
  \bibinfo{year}{2003};\bibinfo{volume}{20}(\bibinfo{number}{2}):\bibinfo{pages}{32--37}.
\newblock \DOIprefix\doi{10.1201/1078/43204.20.2.20030301/41468.6}.
\bibitem[{Singh et~al.(2021)Singh, Gupta, Busso and Kamboj}]{Singh_2021}
\bibinfo{author}{Singh\xfnm[ S.K.]}, \bibinfo{author}{Gupta\xfnm[ S.]},
  \bibinfo{author}{Busso\xfnm[ D.]}, \bibinfo{author}{Kamboj\xfnm[ S.]}.
\newblock \bibinfo{title}{Top management knowledge value, knowledge sharing
  practices, open innovation and organizational performance}.
\newblock \bibinfo{journal}{Journal of Business Research}
  \bibinfo{year}{2021};\bibinfo{volume}{128}:\bibinfo{pages}{788--798}.
\newblock \URLprefix
  \url{https://www.sciencedirect.com/science/article/pii/S0148296319302930}.
  \DOIprefix\doi{https://doi.org/10.1016/j.jbusres.2019.04.040}.
\bibitem[{Sternberg et~al.(2000)Sternberg, Forsythe, Hedlund, Horvath, Wagner,
  Williams, Snook and Grigorenko}]{Sternberg_2000}
\bibinfo{author}{Sternberg\xfnm[ R.J.]}, \bibinfo{author}{Forsythe\xfnm[
  G.B.]}, \bibinfo{author}{Hedlund\xfnm[ J.]}, \bibinfo{author}{Horvath\xfnm[
  J.A.]}, \bibinfo{author}{Wagner\xfnm[ R.K.]}, \bibinfo{author}{Williams\xfnm[
  W.M.]}, \bibinfo{author}{Snook\xfnm[ S.A.]},
  \bibinfo{author}{Grigorenko\xfnm[ E.L.]}.
\newblock \bibinfo{title}{Practical intelligence in everyday life}.
\newblock \bibinfo{publisher}{Cambridge University Press},
  \bibinfo{year}{2000}.
\bibitem[{Szulanski(1996)}]{Szulanski_1996}
\bibinfo{author}{Szulanski\xfnm[ G.]}.
\newblock \bibinfo{title}{Exploring internal stickiness: Impediments to the
  transfer of best practice within the firm}.
\newblock \bibinfo{journal}{Strategic management journal}
  \bibinfo{year}{1996};\bibinfo{volume}{17}(\bibinfo{number}{S2}):\bibinfo{pages}{27--43}.
\newblock \DOIprefix\doi{10.1002/smj.4250171105}.
\bibitem[{Taweel et~al.(2009)Taweel, Delaney, Arvanitis and Zhao}]{Taweel_2009}
\bibinfo{author}{Taweel\xfnm[ A.]}, \bibinfo{author}{Delaney\xfnm[ B.]},
  \bibinfo{author}{Arvanitis\xfnm[ T.N.]}, \bibinfo{author}{Zhao\xfnm[ L.]}.
\newblock \bibinfo{title}{Communication, knowledge and co-ordination management
  in globally distributed software development: Informed by a scientific
  software engineering case study}.
\newblock In: \bibinfo{booktitle}{2009 Fourth IEEE International Conference on
  Global Software Engineering}. \bibinfo{organization}{IEEE};
  \bibinfo{year}{2009}. p. \bibinfo{pages}{370--375}.
\newblock \DOIprefix\doi{10.1109/icgse.2009.58}.
\bibitem[{Trkman and Desouza(2012)}]{Trkman_2012}
\bibinfo{author}{Trkman\xfnm[ P.]}, \bibinfo{author}{Desouza\xfnm[ K.C.]}.
\newblock \bibinfo{title}{Knowledge risks in organizational networks: An
  exploratory framework}.
\newblock \bibinfo{journal}{The Journal of Strategic Information Systems}
  \bibinfo{year}{2012};\bibinfo{volume}{21}(\bibinfo{number}{1}):\bibinfo{pages}{1--17}.
\bibitem[{Voelpel et~al.(2005)Voelpel, Dous and Davenport}]{Voelpel_2005}
\bibinfo{author}{Voelpel\xfnm[ S.C.]}, \bibinfo{author}{Dous\xfnm[ M.]},
  \bibinfo{author}{Davenport\xfnm[ T.H.]}.
\newblock \bibinfo{title}{Five steps to creating a global knowledge-sharing
  system: Siemens' sharenet}.
\newblock \bibinfo{journal}{Academy of Management Perspectives}
  \bibinfo{year}{2005};\bibinfo{volume}{19}(\bibinfo{number}{2}):\bibinfo{pages}{9--23}.
\newblock \DOIprefix\doi{10.5465/ame.2005.16962590}.
\bibitem[{Wang and Noe(2010)}]{Wang_2010}
\bibinfo{author}{Wang\xfnm[ S.]}, \bibinfo{author}{Noe\xfnm[ R.A.]}.
\newblock \bibinfo{title}{Knowledge sharing: A review and directions for future
  research}.
\newblock \bibinfo{journal}{Human Resource Management Review}
  \bibinfo{year}{2010};\bibinfo{volume}{20}(\bibinfo{number}{2}):\bibinfo{pages}{115--131}.
\newblock \DOIprefix\doi{10.1016/j.hrmr.2009.10.001}.
\bibitem[{Wegner(1987)}]{Wegner_1987}
\bibinfo{author}{Wegner\xfnm[ D.M.]}.
\newblock \bibinfo{title}{Transactive memory: A contemporary analysis of the
  group mind}.
\newblock \bibinfo{journal}{Theories of group behavior}
  \bibinfo{year}{1987};:\bibinfo{pages}{185--208}\DOIprefix\doi{10.1007/978-1-4612-4634-3_9}.
\bibitem[{Wendling et~al.(2013)Wendling, Oliveira and Carlos
  Gastaud~Maçada}]{Wendling_2013}
\bibinfo{author}{Wendling\xfnm[ M.]}, \bibinfo{author}{Oliveira\xfnm[ M.]},
  \bibinfo{author}{Carlos Gastaud~Maçada\xfnm[ A.]}.
\newblock \bibinfo{title}{Knowledge sharing barriers in global teams}.
\newblock \bibinfo{journal}{Journal of Systems and Information Technology}
  \bibinfo{year}{2013};\bibinfo{volume}{15}(\bibinfo{number}{3}):\bibinfo{pages}{239--253}.
\newblock \DOIprefix\doi{10.1108/jsit-09-2012-0054}.
\bibitem[{Witherspoon et~al.(2013)Witherspoon, Bergner, Cockrell and
  Stone}]{Witherspoon_2013}
\bibinfo{author}{Witherspoon\xfnm[ C.]}, \bibinfo{author}{Bergner\xfnm[ J.]},
  \bibinfo{author}{Cockrell\xfnm[ C.]}, \bibinfo{author}{Stone\xfnm[ D.]}.
\newblock \bibinfo{title}{Antecedents of organizational knowledge sharing: a
  meta-analysis and critique}.
\newblock \bibinfo{journal}{Journal of Knowledge Management}
  \bibinfo{year}{2013};\DOIprefix\doi{10.1108/13673271311315204}.
\bibitem[{Zieba(2023)}]{Zieba_2023}
\bibinfo{author}{Zieba\xfnm[ M.]}.
\newblock \bibinfo{title}{Emotions and their relation with knowledge risks in
  organizations}.
\newblock In: \bibinfo{booktitle}{The Future of Knowledge Management:
  Reflections from the 10th Anniversary of the International Association of
  Knowledge Management (IAKM)}. \bibinfo{publisher}{Springer};
  \bibinfo{year}{2023}. p. \bibinfo{pages}{169--184}.

\end{thebibliography}

\appendix

\pagebreak
\section{Interview structure}
\label{sec-appendix}

\textbf{Introduction}

\begin{itemize}
    \item Researcher's introduction
    \item Reminder of study's purpose, goals, and scope
    \item Facts about the interview – recording, privacy, data processing
    \item Importance of honesty, policy to not judge
    \item Any questions before the recording starts
\end{itemize}

\textbf{Theme 1: Information/knowledge owned by you and your team}
\begin{itemize}
    \item Types of documents, assets, tools, and resources
    \item Teams and individuals benefit
    \item Importance of information for others
\end{itemize}

\textbf{Theme 2: Information/knowledge needed by you and your team}
\begin{itemize}
    \item Types of documents, assets, tools, and resources
    \item Teams and individuals owning that information
    \item Location and availability
    \item Importance of information for you
\end{itemize}

\textbf{Theme 3: Knowledge sharing and knowledge management}
\begin{itemize}
    \item Importance, benefits, personal motivation
    \item Activities and initiatives in the area
    \item Current state vs. desired state
    \item Challenges
\end{itemize}

\textbf{Theme 4: Recommendations to improve}
\begin{itemize}
    \item Encouragement and motivation
    \item Ability and capability
    \item Responsibility
    \item Concrete ideas for more effective knowledge-sharing and management
\end{itemize}

\end{document}